\documentclass[onecolumn]{aastex62}

\usepackage{subfigure}
\usepackage{tabularx}
\usepackage{multirow}
\usepackage{xcolor}
\usepackage{amsmath,amssymb}

\received{September 21, 2019}
\revised{November 18, 2019}
\accepted{November 28, 2019}

\submitjournal{ApJ}
\shorttitle{The dust gap-ring in the CR Cha protoplanetary disk}
\shortauthors{Kim et al.}

\begin{document}

\title{The detection of dust gap-ring structure in the outer region of the CR Cha protoplanetary disk}


\author[0000-0003-3085-0690]{Seongjoong Kim}
\affil{ Department of Earth and Planetary Sciences, Tokyo Institute of Technology, 
2-12-1 Ookayama, Meguro-ku, Tokyo 152-8551, Japan}

\author[0000-0003-3038-364X]{Sanemichi Takahashi}
\affil{ National Astronomical Observatory of Japan, 2-21-1 Osawa, Mitaka, Tokyo 181-8588, Japan}
\affil{ Department of Applied Physics, Kogakuin University, 1-24-2 Nishi-Shinjuku, Shinjuku-ku, Tokyo, 163-8677, Japan}

\author[0000-0002-7058-7682]{Hideko Nomura}
\affil{ National Astronomical Observatory of Japan, 2-21-1 Osawa, Mitaka, Tokyo 181-8588, Japan}
\affil{ Department of Earth and Planetary Sciences, Tokyo Institute of Technology,
2-12-1 Ookayama, Meguro-ku, Tokyo 152-8551, Japan}

\author[0000-0002-6034-2892]{Takashi Tsukagoshi}
\affil{ National Astronomical Observatory of Japan, 2-21-1 Osawa, Mitaka, Tokyo 181-8588, Japan}

\author{Seokho Lee}
\affil{ National Astronomical Observatory of Japan, 2-21-1 Osawa, Mitaka, Tokyo 181-8588, Japan}

\author{Takayuki Muto}
\affil{ Division of Liberal Arts, Kogakuin University, 1-24-2 Nishi-Shinjuku, Shinjuku-ku, Tokyo 163-8677, Japan}

\author[0000-0001-9290-7846]{Ruobing Dong}
\affil{Department of Physics \& Astronomy, University of Victoria, Victoria, BC, V8P 1A1, Canada}

\author{Yasuhiro Hasegawa}
\affil{ Jet Propulsion Laboratory, California Institute of Technology, Pasadena, CA 91109, USA}

\author[0000-0002-3053-3575]{Jun Hashimoto}
\affil{ Astrobiology Center, National Institutes of Natural Sciences, 2-21-1 Osawa, Mitaka, Tokyo 181-8588, Japan}

\author[0000-0001-7235-2417]{Kazuhiro Kanagawa}
\affil{Research Center for the Early Universe, Graduate School of Science, University of Tokyo, Hongo, Bunkyo-ku, Tokyo 113-0033, Japan }

\author[0000-0003-4562-4119]{Akimasa Kataoka}
\affil{  National Astronomical Observatory of Japan, 2-21-1 Osawa, Mitaka, Tokyo 181-8588, Japan}

\author[0000-0003-0114-0542]{Mihoko Konishi}
\affil{ Faculty of Science and Technology, Oita University, 700 Dannoharu, Oita 870-1192, Japan}

\author[0000-0003-2300-2626]{Hauyu Baobab Liu}
\affil{ Academia Sinica Institute of Astronomy and Astrophysics, P.O. Box 23-141, Taipei 10617, Taiwan}

\author{Munetake Momose}
\affil{College of Science, Ibaraki University, 2-1-1 Bunkyo, Mito, Ibaraki 310-8512, Japan}

\author[0000-0003-1799-1755]{Michael Sitko}
\affil{ Department of Physics, University of Cincinnati, Cincinnati, OH 45221, USA}
\affil{ Space Science Institute, 475 Walnut Street, Suite 205, Boulder, CO 80301, USA}

\author[0000-0001-8105-8113]{Kengo Tomida}
\affil{ Department of Earth and Space Science, Osaka University, Toyonaka, Osaka 560-0043, Japan }

\begin{abstract}

We observe the dust continuum at 225 GHz and CO isotopologue ($^{12}$CO, $^{13}$CO, and C$^{18}$O) $J=2-1$ emission lines toward the CR Cha protoplanetary disk using the Atacama Large Millimeter/submillimeter Array (ALMA). 
The dust continuum image shows a dust gap-ring structure in the outer region of the dust disk. 
A faint dust ring is also detected around 120 au beyond the dust gap. 
The CO isotopologue lines indicate that the gas disk is more extended than the dust disk. The peak brightness temperature of the $^{13}$CO line shows a small bump around 130 au while $^{12}$CO and C$^{18}$O lines do not show.
We investigate two possible mechanisms for reproducing the observed dust gap-ring structure and a gas temperature bump. First, the observed gap structure can be opened by a Jupiter mass planet using the relation between the planet mass and the gap depth and width. Meanwhile, the radiative transfer
calculations based on the observed dust surface density profile 
show that the observed dust ring could be formed by dust accumulation at the gas temperature bump, that is, the gas pressure bump produced beyond the outer edge of the dust disk.

\end{abstract}


\section{Introduction} 
\label{sec:intro}

CR Cha (a.k.a Sz 6, Cha T8) is a K-type pre-main sequence star \citep{Hussain2009,Villebrun2019} in Cha I dark cloud, one of the famous star forming clouds in our galaxy. 
The fitting on the stellar evolution tracks suggests that the stellar mass of CR Cha is $M_{\star}= 1.2-2 \rm~M_{\odot}$ with an age of $1-3$ Myr \citep[e.g.,][]{DAntona1994,Natta2000,Hussain2009,Varga2018}. The latest updated distance of CR Cha is $ d\sim187.5\pm0.8\rm~pc$ \citep[GAIA DR 2;][]{GAIA2018}.
The existence of the protoplanetary disk around CR Cha had already been revealed by the thermal dust emission at (sub-)mm wavelengths \citep[e.g.,][]{Ubach2012,Pascucci2016,Ribas2017,Ubach2017}. 

The spectral index $\alpha$ at mm-wavelength is linked to the dust properties, such as dust opacity and the dust grain size \cite[e.g.,][]{Miyake1993}. It can be used to study the dust growth in the disk by comparing with the spectral index of the interstellar medium (ISM) \citep[e.g.,][]{Beckwith1991,DAlessio2001}. 
Toward the CR Cha disk, the measured spectral index at 1 mm is $\alpha_{\rm 1mm}\sim3.4$ \citep{Ubach2012}, which is close to $\alpha_{\rm ISM}\sim3.7$ in the ISM. 
This large $\alpha_{\rm 1mm}$ value in the CR Cha disk indicates that the maximum grain size is small in the outer disk \citep[e.g.,][]{Draine2006}.
This may be the results of the dust growth in this region: the mm-sized dust grains already drifted inward in the gas disk due to the gas drag \citep{Ribas2017,Ubach2017}.
Recent CO line survey toward the protoplanetary disks in Cha I cloud using Atacama Large Millimeter/submillimeter Array (ALMA) can support this possibility by detecting that the gas disk around the CR Cha exists with an angular size of $\sim0.9''$ in radius \citep{Long2017}. 

However, the time scale of the radial drift of mm/cm-sized dust grains \citep[$10^{4-5}$ yr; e.g.,][]{Takeuchi2005,Brauer2007} is much shorter than the age of CR Cha ($\sim1-3~\rm Myr$).
\cite{Ribas2017} suggested that some braking mechanisms, such as dust trapping at the pressure bumps produced by zonal flow \citep[e.g.,][]{Pinilla2012} or a planet-opened gaps \citep[e.g.,][]{Zhu2012}, are required to stop or slow down the radial drift of the dust grains. 
That is, the CR Cha disk is one of the strong candidates which would have disk substructures like a dust ring/gap.

In this paper, we show the results of the dust continuum and the CO isotopologue $J=2-1$ emission lines observed at ALMA Band 6 to examine the existence of small scale disk substructure like a gap or ring structure in the CR Cha disk.
We summarize the information of ALMA observations toward the CR Cha disk in Section \ref{sec:observation}. The observed dust continuum and CO isotopologue images are presented in Section \ref{sec:result}. In Section \ref{sec:discussion}, we discuss the radial profiles of dust surface density and CO gas column density derived from the observations and the possible mechanisms to explain the detected CR Cha disk structure. Finally, we summarize our results in Section \ref{sec:summary}.

\begin{table*}[p]
\begin{center}
\caption{The summary of dust continuum and CO isotopologue line images.}
\begin{tabular}{c|c|c|c|c|c}
\hline \hline \noalign {\smallskip}
Species & $\nu_{\rm center}$ & RMS in Cube & RMS in Mom. 0 & Beam size & Beam PA  \\
 & [GHz] & [mJy beam$^{-1}$] & [mJy beam$^{-1}$ km s$^{-1}$] & [arcsec$^{2}$] & [$^{\circ}$]  \\
\hline \noalign {\smallskip}
Dust & 225 & $\sim0.0178$ &  & $0.087\times0.052$ & 35.75  \\
$^{12}$CO & 230.538 & $\sim2.47$ & $\sim$4.594 & $0.188\times0.128$ & 0.635 \\
$^{13}$CO & 220.398 & $\sim2.40$ & $\sim$4.978 & $0.196\times0.143$ & 0.405 \\
C$^{18}$O & 219.560 & $\sim2.28$ & $\sim$3.441 & $0.194\times0.142$ & 0.976 \\ 
\hline \noalign {\smallskip}
\end{tabular}
\end{center}
\label{tab:table1}
\end{table*}

\begin{figure*}[p]
\centering  
\subfiguretopcaptrue
\subfigure[Dust continuum at 225 GHz]{\label{fig:1a}\includegraphics[width=0.45\textwidth]{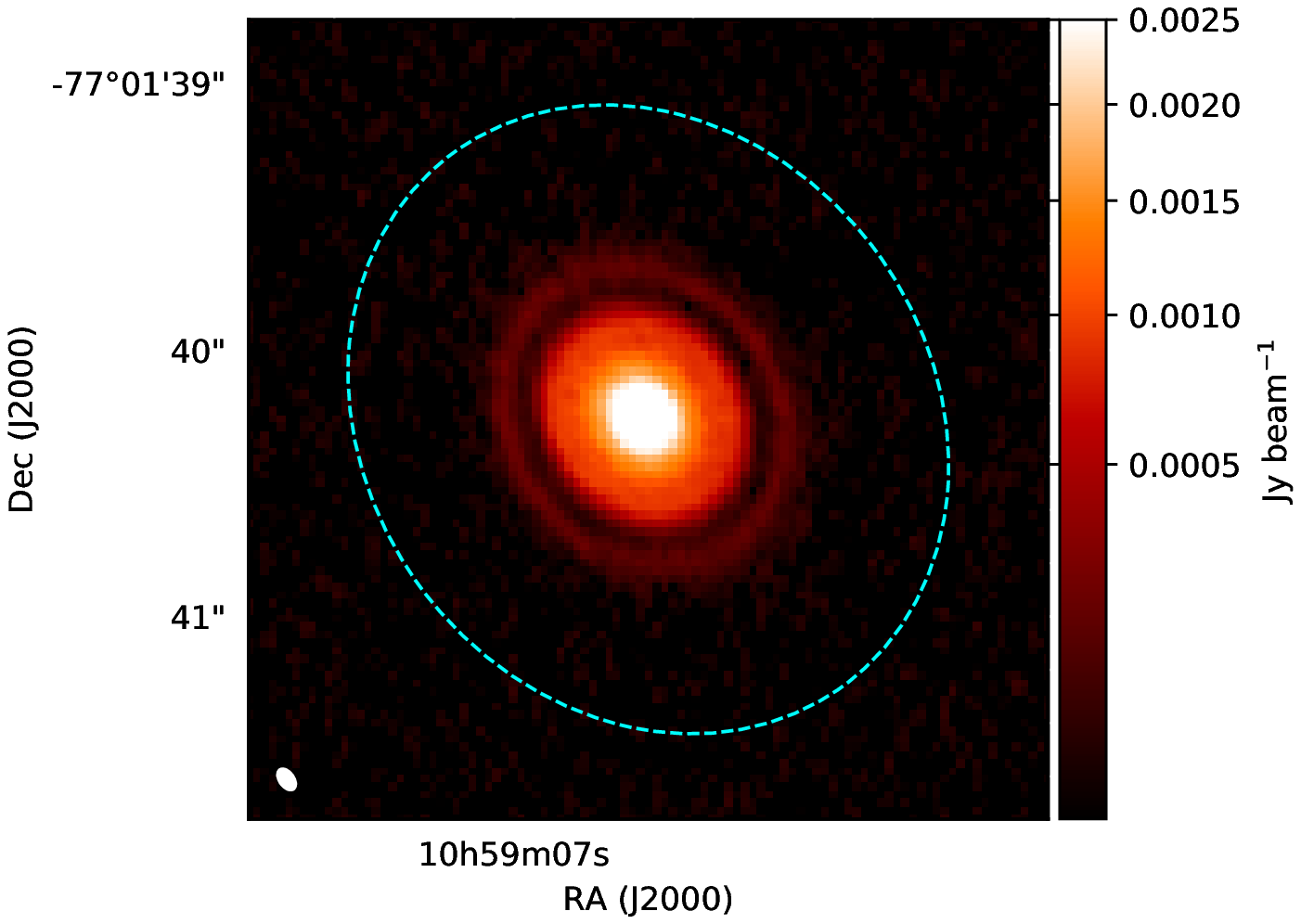}}
\subfigure[$^{12}$CO emission line]{\label{fig:1b}\includegraphics[width=0.45\textwidth]{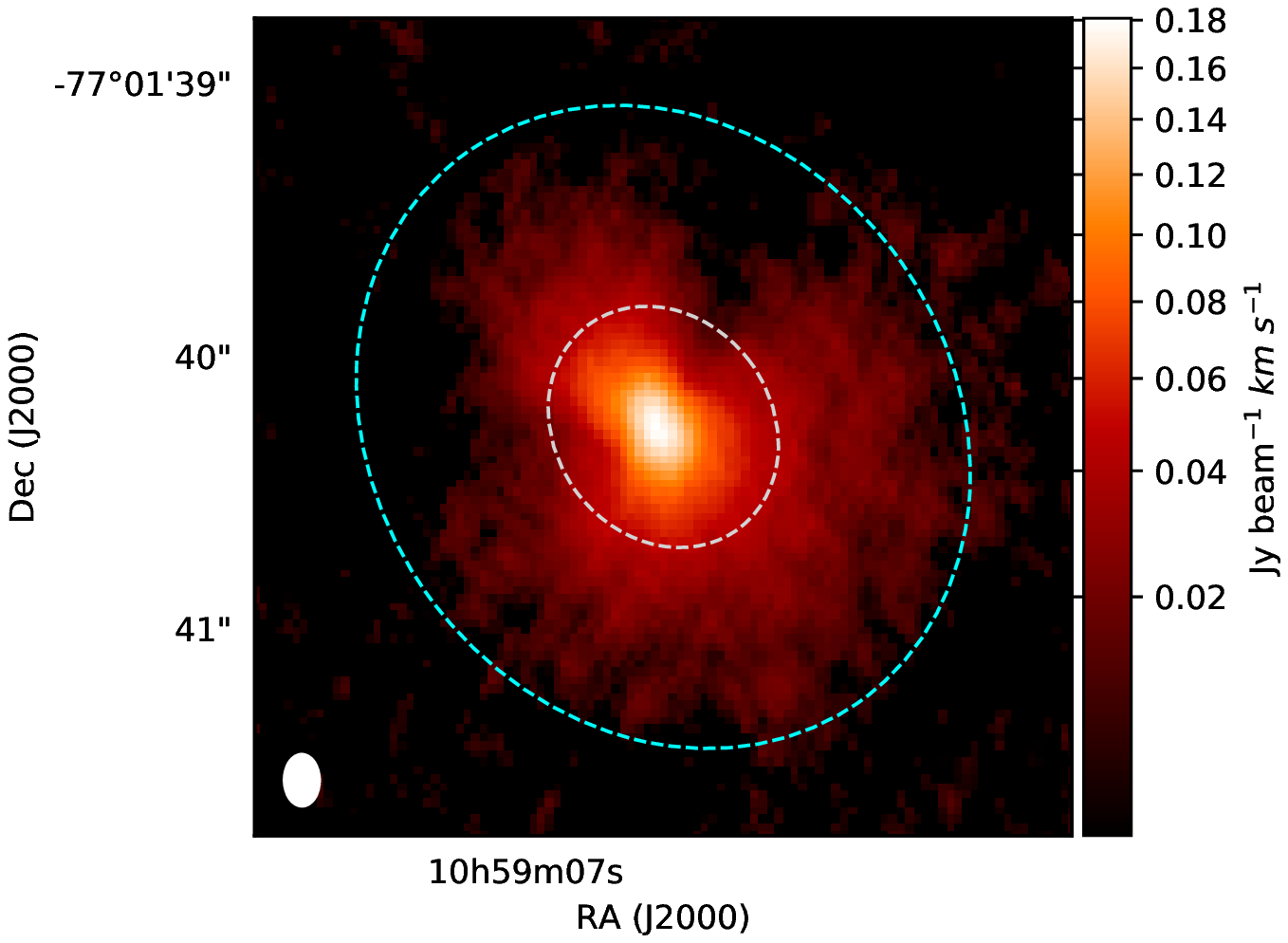}}
\hfill \subfigure[$^{13}$CO emission line]{\label{fig:1c}\includegraphics[width=0.45\textwidth]{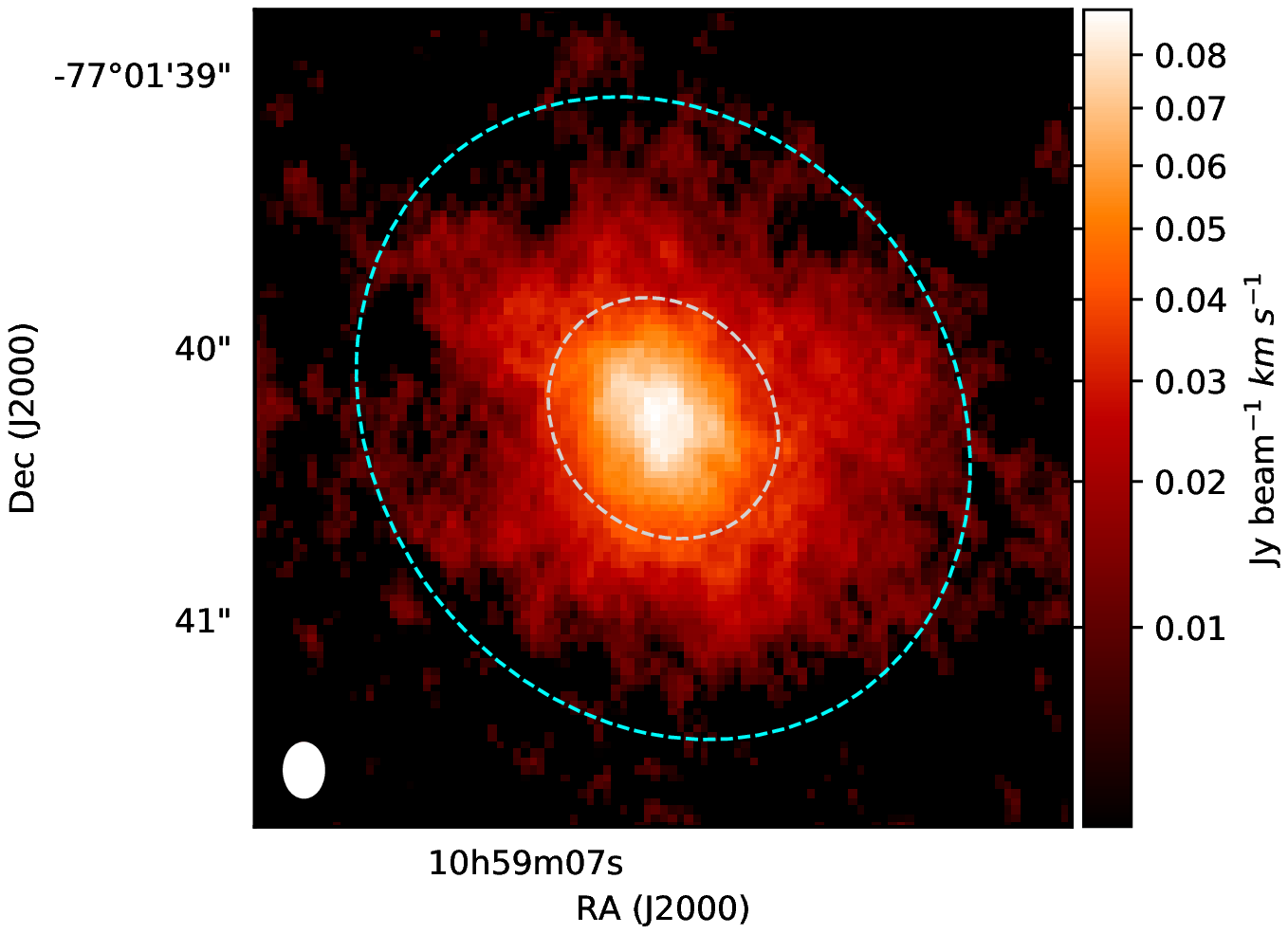}}
\subfigure[C$^{18}$O emission line]{\label{fig:1d}\includegraphics[width=0.45\textwidth]{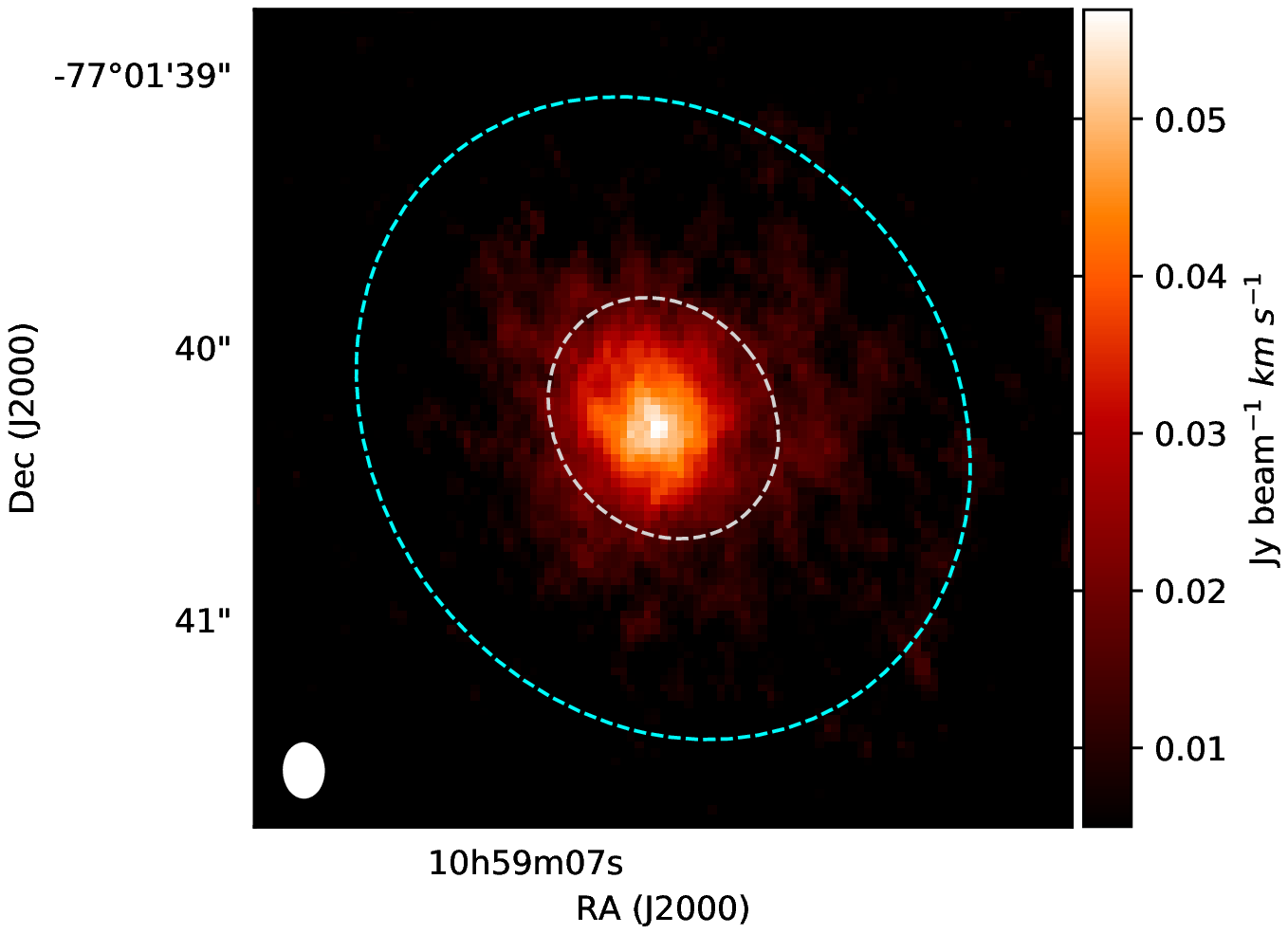}}
\subfiguretopcapfalse
\caption{(a) The 225 GHz dust continuum image and the moment 0 maps of (b) $^{12}$CO, (c) $^{13}$CO, and (d) C$^{18}$O $J=2-1$ line emission of the CR Cha disk. The synthesized beam is presented in the bottom-left corner: $\sim0.087''\times0.052''$ for the dust continuum emission (Briggs robust=0), $\sim0.188''\times0.128''$ for $^{12}$CO, $\sim0.196''\times0.143''$ for $^{13}$CO, and $\sim0.194''\times0.142''$ for C$^{18}$O line emission (Briggs robust=2). The colorbar indicates the observed intensity in $\rm Jy~beam^{-1}$ unit for the dust continuum and the integrated intensities in $\rm Jy~beam^{-1}~km~s^{-1}$ unit for the CO isotopologue lines. 
The white dashed ellipse shows the location of the dust gap at $\rm r=90~au$. A faint dust ring is detected beyond the dust gap. The cyan dashed ellipse indicates the outer edge of the CO gas disks at $\rm r=240~au$.}
\label{fig:figure1}
\end{figure*}

\section{Observation} 
\label{sec:observation}

The CR Cha protoplanetary disk is observed at ALMA Band 6 on November 2017 and March 2018 with 47 and 42 antennae respectively (2017.1.00286.S). The projected baselines are from 12.54 m to 7.908 km, corresponding to a maximum angular coverage of $\sim5''$ and an angular resolution of $\gtrsim0.06''$. The total execution time for the observations is $\sim3.2$ hours including the target on-source time of $\sim2$ hours. J1427-4206 and J0635-7516 are used for the bandpass and flux calibrations at each observation. J1058-8003 is selected as a phase calibrator at both observations.

The correlator setup has four spectral windows (SPWs). Dust continuum emission is observed at 233 GHz (SPW1) and 217 GHz (SPW2) with 1.8 GHz bandwidth at each SPW. For the CO isotopologue emission lines, the SPWs are centered at 230.538 GHz for $^{12}$CO (SPW0), 220.398 GHz for $^{13}$CO, and 219.560 GHz for C$^{18}$O (SPW3). The spectral resolution for three CO isotopologue lines is 61 kHz, corresponding to $\rm \sim0.08~km~s^{-1}$ in velocity.   

The basic calibration steps are progressed through the ALMA calibration pipeline using \texttt{CASA} package version 5.3. After the ALMA pipeline calibration, we make the images of dust continuum and CO isotopologue emission lines by \texttt{CLEAN} command in \texttt{CASA}. The Briggs $\rm robust=0$ is used for the dust continuum image, obtaining the synthesized beam size of $\sim0.087''\times0.052''$ and the beam position angle (PA) of $\sim35.75^{\circ}$. The phase self-calibration is applied to the dust continuum image for improving the data quality. Meanwhile, the Briggs $\rm robust=2.0$ (Natural weighting) is applied for the CO isotopologue lines to obtain high signal-to-noise ratio $\rm (SNR)\sim20-30$. We also apply the phase self-calibration to the CO line data but the images are not improved. We note that the SNRs of the CO isotopologue lines are $\sim5-7$ with $\sim0.080''\times0.047''$ synthesized beam size when the $\rm robust=0$ is applied. The synthesized beam size is $\sim0.188''\times0.128''$ for the $^{12}$CO line, $\sim0.196''\times0.143''$ for the $^{13}$CO line, and $\sim0.194''\times0.142''$ for C$^{18}$O line, respectively. Table \ref{tab:table1} summarizes the angular resolution and RMS noise level of the observed data. 

\begin{figure*}[p]
\centering  
\subfiguretopcaptrue
\mbox{}\hfill
\subfigure[$^{12}$CO spectrum]{\label{fig:3a}\includegraphics[width=0.325\textwidth]{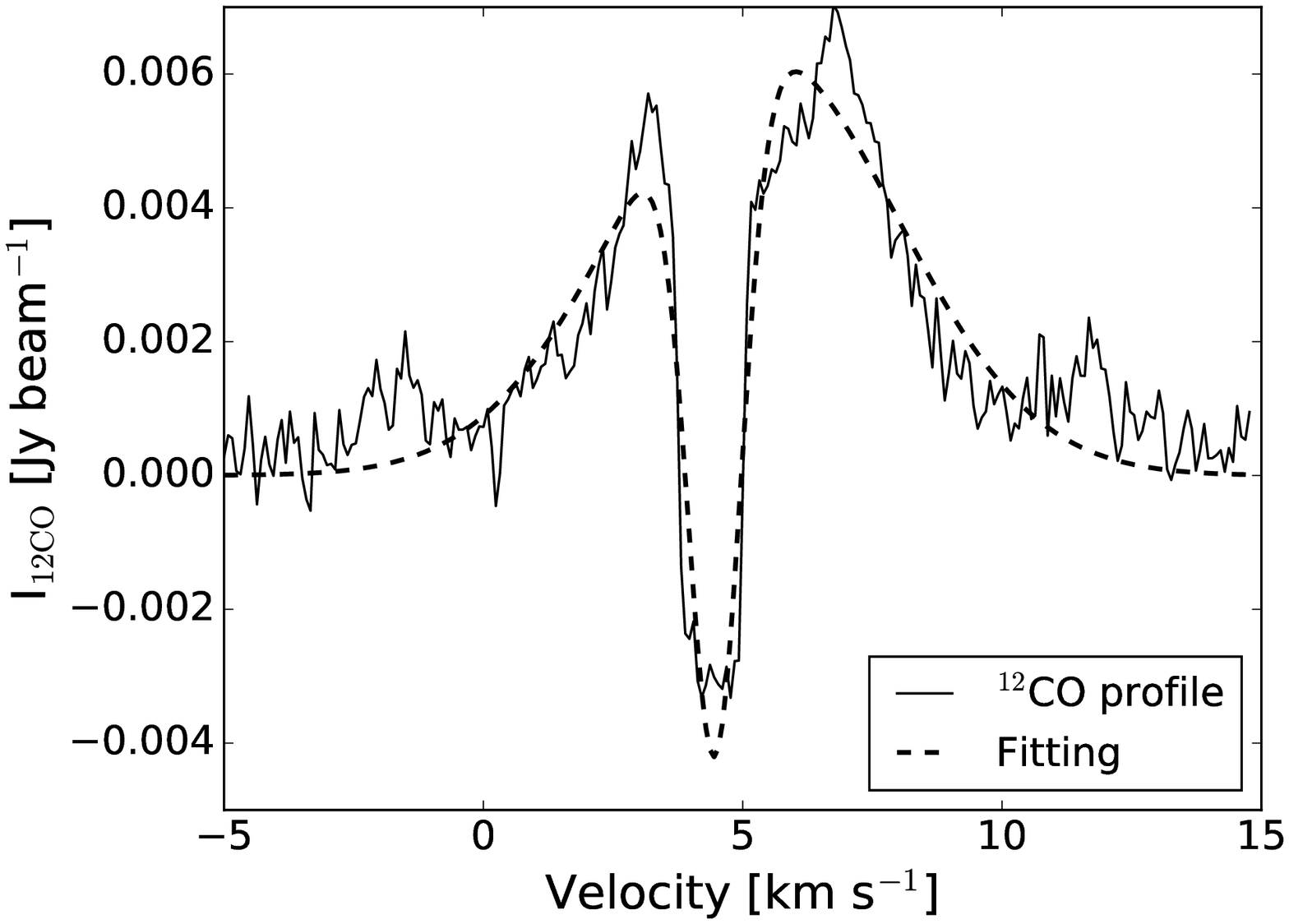}}
\subfigure[$^{13}$CO spectrum]{\label{fig:3b}\includegraphics[width=0.325\textwidth]{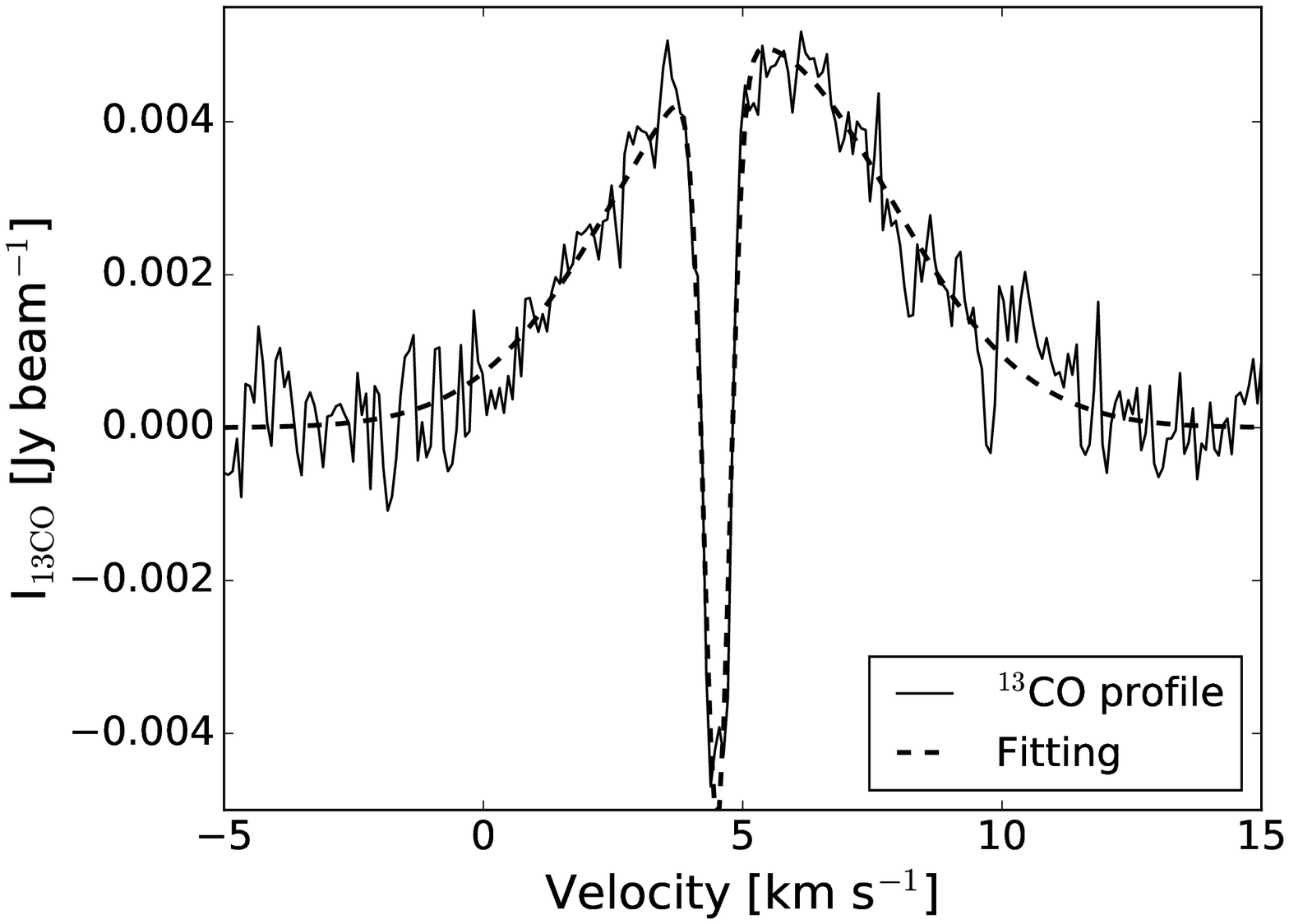}}
\subfigure[C$^{18}$O spectrum]{\label{fig:3c}\includegraphics[width=0.325\textwidth]{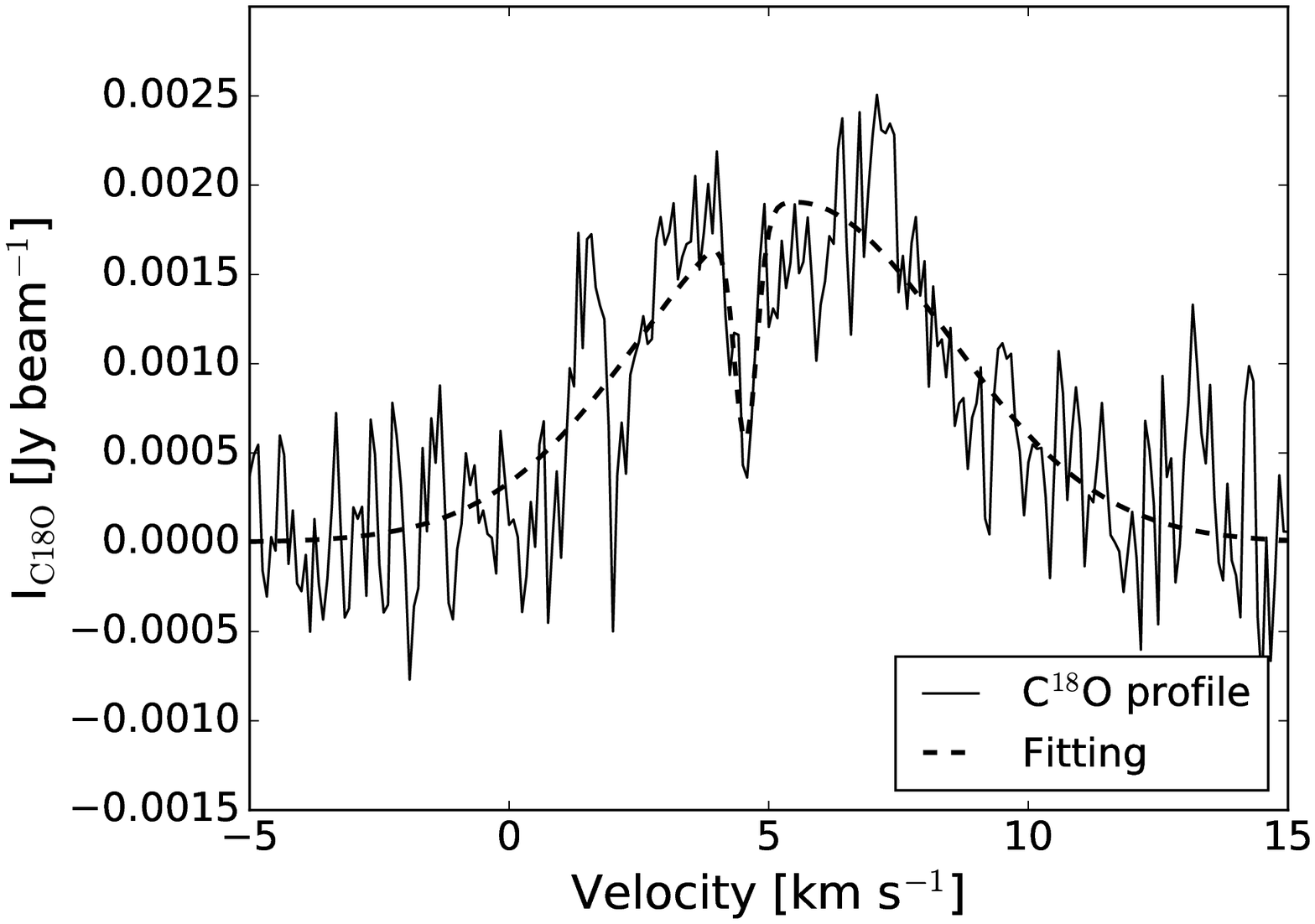}}
\hfill\mbox{}
\subfiguretopcapfalse
\caption{The averaged spectra (solid line) of (a) $^{12}$CO, (b) $^{13}$CO, and (c) C$^{18}$O $J=2-1$ emission lines inside the disk radius of $\rm r\lesssim240$ au. The dashed lines are the Gaussian fitting of the emission plus absorption profiles. The emission lines are centered at $\rm \sim5.3~km~s^{-1}$ with $\Delta V_{\rm emis}$ of $\rm \sim2.7~km~s^{-1}$ on average. The absorption features are centered at $\rm \sim4.5~km~s^{-1}$ with $\Delta V_{\rm abs}$ of $\rm \sim0.2-0.5~km~s^{-1}$. The fitting parameters are summarized in Table \ref{tab:table2}.}
\label{fig:figure3}
\end{figure*}

\begin{table*}[p]
\begin{center}
\caption{The summary of Gaussian fitting parameters for the emission and absorption of the CO isotopologue lines.}
\begin{tabular}{c|c|c|c|c|c|c}
\hline \hline \noalign {\smallskip}
Species & $A_{\rm emis}$ & $V_{\rm emis}$ & $\Delta V_{\rm emis}$ &  $A_{\rm abs}$ & $V_{\rm abs}$ & $\Delta V_{\rm abs}$  \\
 & [mJy beam$^{-1}$] & [km s$^{-1}$] & [km s$^{-1}$] &   [mJy beam$^{-1}$] & [km s$^{-1}$] & [km s$^{-1}$]  \\
\hline \noalign {\smallskip}
$^{12}$CO & 6.37$\pm$0.22 & 5.30$\pm$0.07 & 2.67$\pm$0.07 & -10.26$\pm$0.34 & 4.46$\pm$0.02 & 0.52$\pm$0.02 \\
$^{13}$CO & 5.00$\pm$0.11 & 5.19$\pm$0.06 & 2.64$\pm$0.06 & -9.97$\pm$0.30 & 4.50$\pm$0.01 & 0.25$\pm$0.01 \\
C$^{18}$O & 1.90$\pm$0.08 & 5.51$\pm$0.13 & 2.96$\pm$0.13 & -1.24$\pm$0.26 & 4.56$\pm$0.05 & 0.21$\pm$0.05 \\ 
\hline \noalign {\smallskip}
\end{tabular}
\end{center}
\label{tab:table2}
\end{table*}

\begin{figure*}[p]
\centering  
\subfiguretopcaptrue
\hfill
\subfigure[The azimuthally averaged (integrated) intensity profiles]{\label{fig:2b}\includegraphics[width=0.48\textwidth]{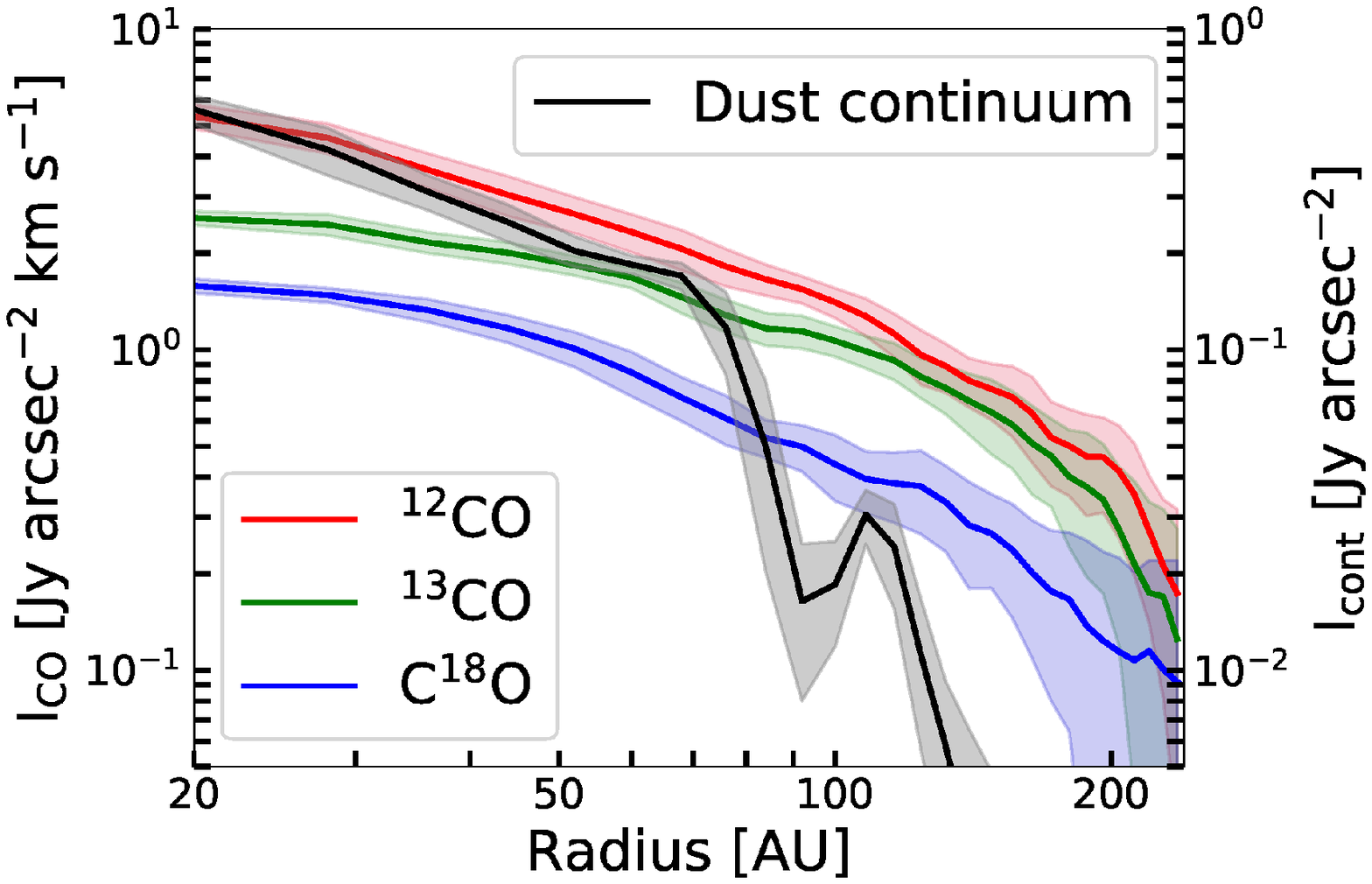}} \hfill
\subfigure[The ratios between the CO isotopologue lines]{\label{fig:2c}\includegraphics[width=0.4\textwidth]{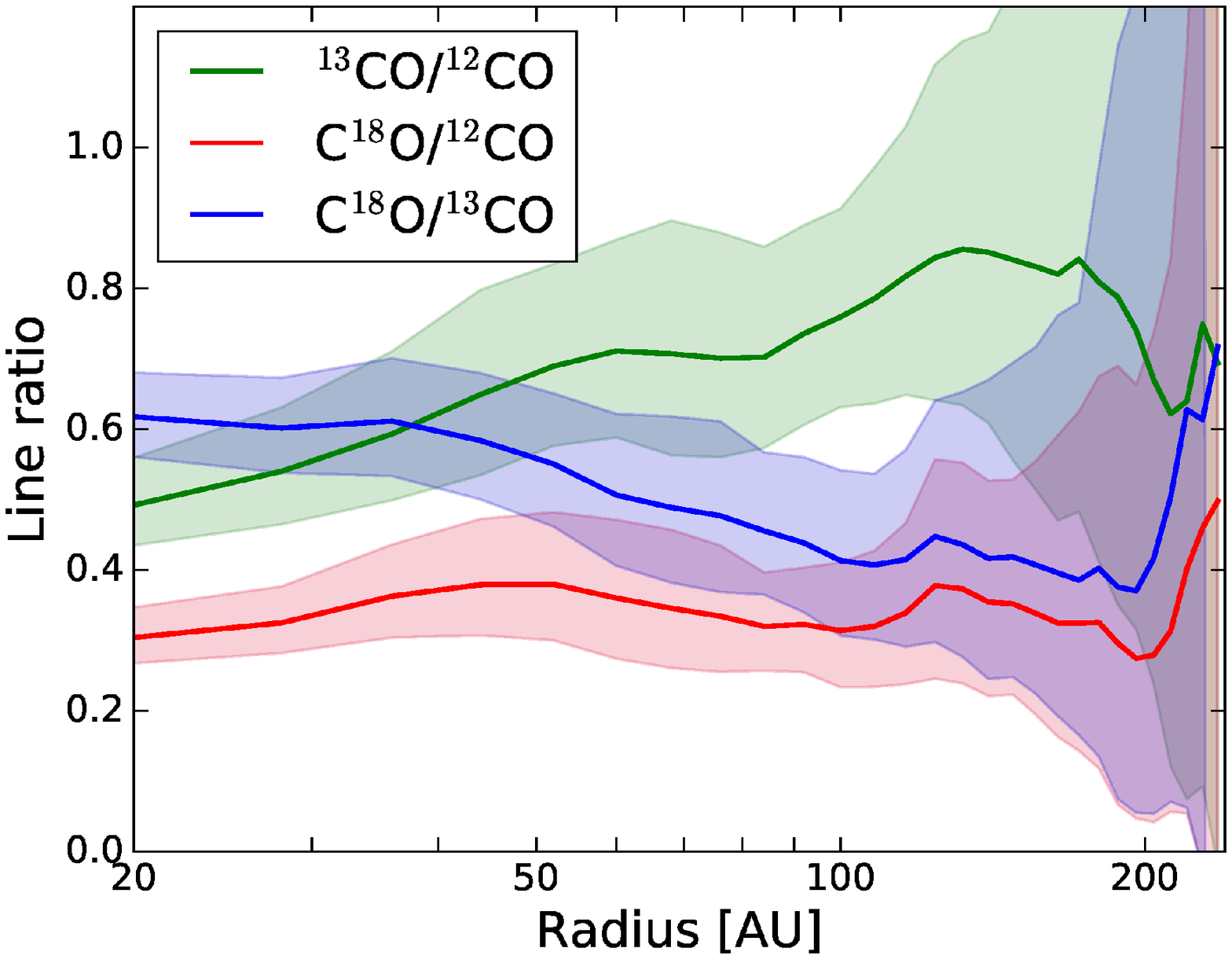}}
\hfill
\subfiguretopcapfalse
\caption{ (a) The azimuthally averaged radial profiles of the dust continuum emission (black) in $\rm Jy~arcsec^{-2}$ unit and the integrated intensities of $^{12}$CO (red), $^{13}$CO (green), and C$^{18}$O (blue) emission lines in $\rm Jy~arcsec^{-2}~km~s^{-1}$ unit. The color-shaded regions indicate the uncertainty due to the azimuthal average. The dust continuum emission shows a gap at $\sim90$ au and a ring at $\sim120$ au. Meanwhile, the CO isotopologue emissions extend to $\rm r\sim240$ au smoothly.
(b) The ratios between the azimuthally averaged integrated intensity profiles of the CO isotopologue lines: $^{13}$CO/$^{12}$CO (green), C$^{18}$O/$^{12}$CO (red), and C$^{18}$O/$^{13}$CO (blue). The color-shaded regions show the uncertainty due to those of azimuthally averaged integrated CO intensity profiles. 
 }
\label{fig:figure2}
\end{figure*}

\section{Result}
\label{sec:result}

\subsection{The images of dust continuum and CO isotopologue emission lines}
\label{subsec:imaging}

Figure \ref{fig:figure1}(a) shows the dust continuum image at 225 GHz. 
There is a dust gap at $\rm r\sim90$ au and a faint dust ring just beyond the dust gap. The total flux density of the dust disk is $\sim0.139$ Jy. It is consistent with the previous observation of $F\sim0.125$ Jy at 1.2 mm continuum \citep{Ubach2012}. The $1\sigma$ RMS level of dust continuum image is $\rm \sim0.0178~mJy~beam^{-1}$. The peak intensity of the dust ring is $\sim10\sigma$ RMS level on average. We obtain the disk geometry by the ellipse fitting using \texttt{imfit} command in \texttt{CASA} package. The estimated inclination of the disk is $31.0^{\circ}\pm1.4^{\circ}$ and disk PA is $36.2^{\circ}\pm1.8^{\circ}$. 

\begin{figure*}[t]
\centering  
\subfiguretopcaptrue
\subfigure[The moment 8 map of $^{12}$CO line emission]{\label{fig:4a}\includegraphics[width=0.45\textwidth]{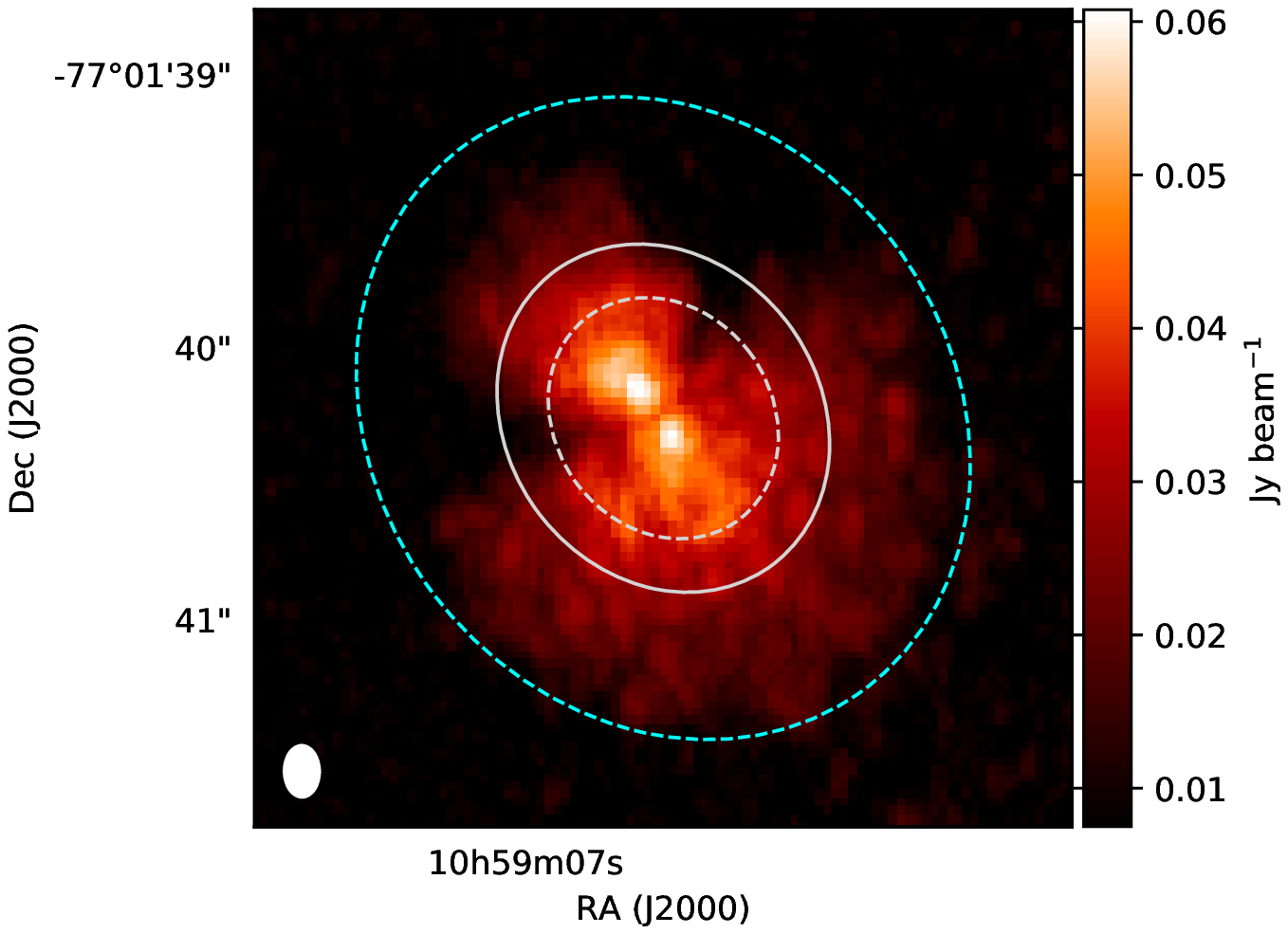}}
\subfigure[The moment 8 map of $^{13}$CO line emission]{\label{fig:4b}\includegraphics[width=0.45\textwidth]{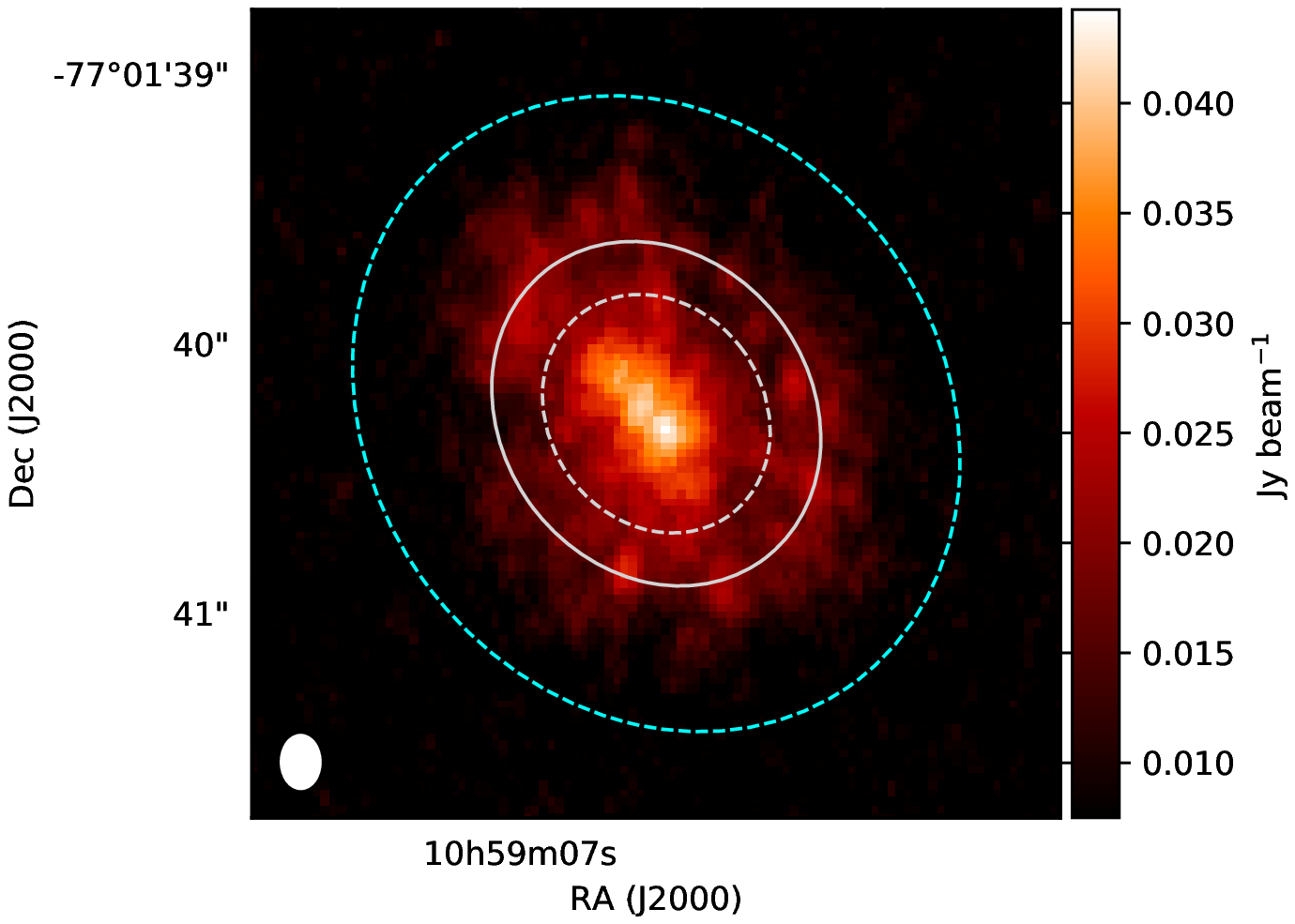}}
\hfill
\subfigure[The azimuthally averaged I$_{\rm \nu,peak}$ profile]{\label{fig:4c}\includegraphics[width=0.45\textwidth]{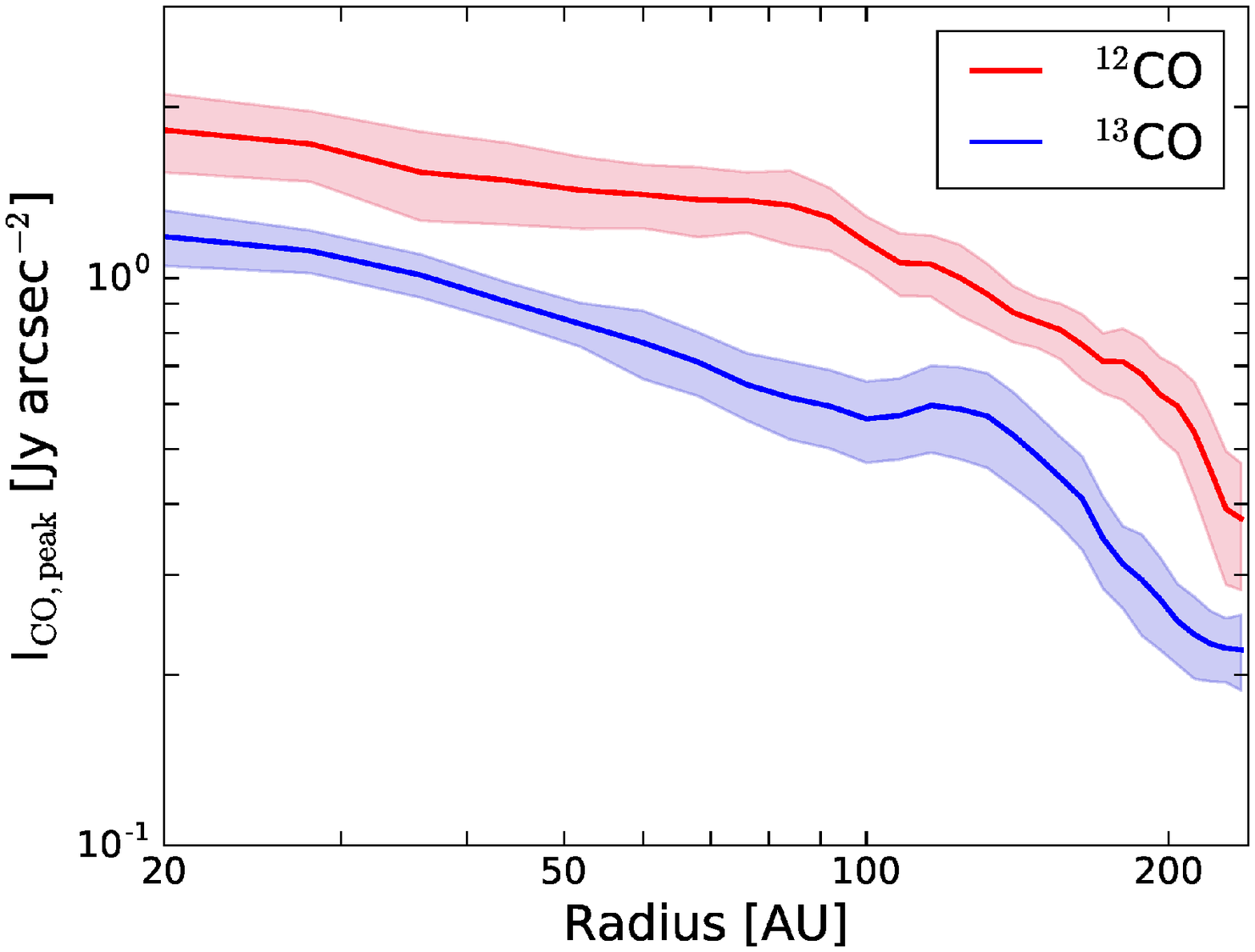}}
\subfigure[The T$_{\rm B,peak}$ profile]{\label{fig:4d}\includegraphics[width=0.45\textwidth]{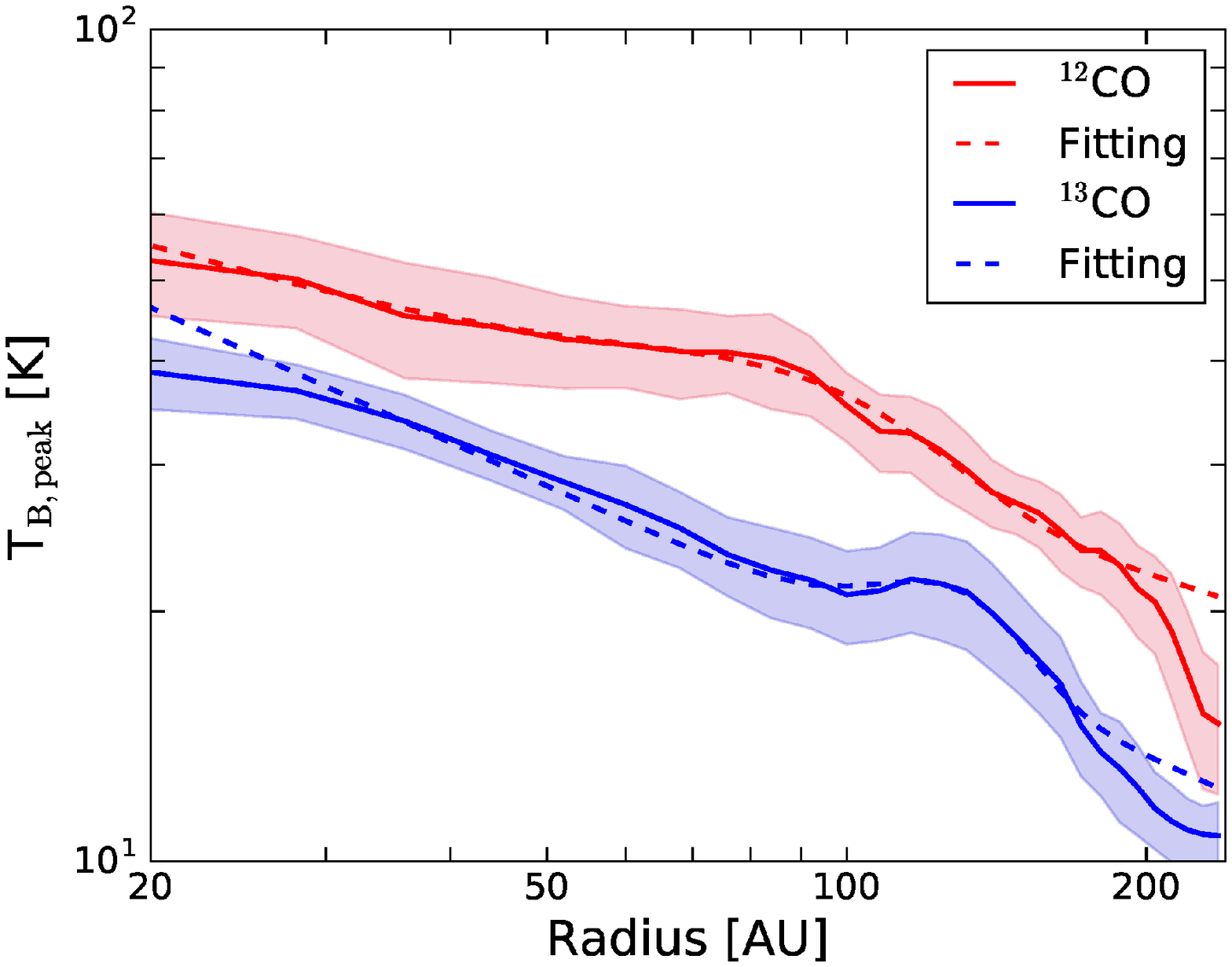}}
\subfiguretopcapfalse
\caption{The moment 8 maps of (a) $^{12}$CO and (b) $^{13}$CO line emission with dust continuum. The absorption feature is seen in the northern half-disk. The overlaid white and cyan dashed ellipses are the same as those in Figure \ref{fig:figure1}. The white solid contour indicates the location of the bump at 130 au in the azimuthally averaged $^{13}$CO peak intensity radial profile.
(c) The azimuthally averaged radial profiles of the peak intensity (I$_{\rm \nu,peak}$) of $^{12}$CO (red) and $^{13}$CO (blue) lines in the southern half-disk which is not affected by the absorption features. The color-shaded regions indicate the uncertainties due to the azimuthal average. There is a small bump in I$_{\rm ^{13}CO,peak}$ around 130 au.
(d) The radial profiles of the peak brightness temperature (T$_{\rm B,peak}$) derived from the I$_{\rm \nu,peak}$ of $^{12}$CO (red) and $^{13}$CO (blue) lines. The color-shaded regions indicate the uncertainties propagated from those of I$_{\rm \nu,peak}$. The dashed lines present the lease square fitting results with Equation \ref{eq:eq21}. The derived T$_{\rm B,peak}$ from the I$_{\rm ^{13}CO,peak}$ profile also shows a small bump at $\rm r\sim130$ au.}
\label{fig:figure4}
\end{figure*}

Figure \ref{fig:figure1}(b), (c), and (d) show the integrated intensity maps (moment 0 maps) of the continuum subtracted line emissions of $^{12}$CO, $^{13}$CO, and C$^{18}$O, respectively. The white dashed ellipse presents the annulus at $\rm r=90~au$ which is corresponding to the dust gap in Figure \ref{fig:figure1}(a). The cyan dashed ellipse indicates the outer edge of the CO gas disk at $\rm r=240~au$. This radius corresponds to the detection limit of the CO isotopologue lines, where the peak intensity at each pixel (the moment 8 map; Figure \ref{fig:figure4}(a) and (b)) is $\rm\sim7.5~mJy~beam^{-1}$, $\approx3\sigma$ RMS level of the data cubes of the CO isotopologue lines. They show that the gas disk of CR Cha is much extended than the dust disk. 

The solid lines in Figure \ref{fig:figure3} present the averaged spectra of (a) $^{12}$CO, (b) $^{13}$CO, and (c) C$^{18}$O line emission inside the disk radius of $\rm r\lesssim240$ au, applying the disk inclination of $31^{\circ}$ and the disk PA of $36.2^{\circ}$. There is a strong absorption feature at the velocity of $\sim4.5$ km s$^{-1}$ in the $^{12}$CO and $^{13}$CO spectra. The absorption feature in C$^{18}$O line spectrum is less prominent than those of $^{12}$CO and $^{13}$CO. 
This absorption feature is also well seen in the channel maps (see Appendix \ref{apsec:channel}) and the moment 8 maps of $^{12}$CO and $^{13}$CO lines (see Figure \ref{fig:figure4}(a) and (b)). 

We perform the least mean square fitting for the emission and absorption lines to measure their strengths, velocity centers, and width. We use the following function for the fitting: 
\begin{equation}
\begin{split}
{\rm I}_{\nu} =  & A_{\rm emit} \exp\left(- \frac{ \left( V-V_{\rm emis}\right)^{2}}{ 2 \Delta V_{\rm emis}^{2}} \right) \\
 & + A_{\rm abs} \exp\left(- \frac{\left( V-V_{\rm abs}\right)^{2}}{ 2 \Delta V_{\rm abs}^{2}} \right).
\end{split}
\end{equation}
The fitting parameters are summarized in Table \ref{tab:table2}. The dashed lines in Figure \ref{fig:figure3} show the fitting results of the emission plus absorption profiles. The fitting results show that the CO emission lines are centered at $\sim5.3$ km s$^{-1}$ with $\Delta V_{\rm emis}$ of $\sim2.7$ km s$^{-1}$ on average and the absorption feature is centered at $\sim4.5$ km s$^{-1}$ with $\Delta V_{\rm abs}$ of $\sim0.2-0.5$ km s$^{-1}$. These absorption features may occur due to the foreground Cha I molecular clouds with a system velocity of $\rm \sim4-5~km~s^{-1}$ \citep[e.g.,][]{Haikala2005,Long2017}. 

\begin{table*}[t]
\begin{center}
\caption{The fitting parameters of the peak brightness temperature profiles of the $^{12}$CO and $^{13}$CO lines.}
\begin{tabular}{c|c|c|c|c|c}
\hline \hline \noalign {\smallskip}
Species & $A$ & $B$ & $C$ & $D$ & $E$  \\
 & [K] &  & [K] & [au] & [au]  \\
\hline \noalign {\smallskip}
$^{12}$CO & 37.57$\pm$1.38 & $-$0.38$\pm$0.02 & 8.37$\pm$0.91 & 82.81$\pm$4.65 & 36.05$\pm$5.69 \\
$^{13}$CO & 28.29$\pm$0.38 & $-$0.54$\pm$0.03 & 4.31$\pm$0.69 & 127.62$\pm$3.33 & 21.96$\pm$4.86 \\
\hline \noalign {\smallskip}
\end{tabular}
\end{center}
\label{tab:table3}
\end{table*}

\subsection{The azimuthally averaged intensity profiles}
\label{subsec:radial_profiles}

To investigate the radial structure of dust and gas disks, we make the azimuthally averaged radial profiles of the images in Figure \ref{fig:figure1}. The derived disk inclination of $31^{\circ}$ and disk PA of $36.2^{\circ}$ are applied to calculate the projected disk radius of each pixel. We azimuthally average every 8 au width annulus inside $\rm r\lesssim240$ au. As mentioned in Section \ref{subsec:imaging}, the CO isotopologue lines are affected by the absorption feature in the northern half-disk against the disk minor axis. We azimuthally average only the southern half-disk (PA = 126.2$^{\circ}$ to 306.2$^{\circ}$, counterclockwise from the North) for the CO isotopologue lines to avoid the absorption feature. Meanwhile, since the dust continuum is free to line absorption feature, we average the full azimuthal angles for the dust continuum image.

Figure \ref{fig:figure2}(a) presents the azimuthally averaged radial profile of dust continuum (black) in the $\rm Jy~arcsec^{-2}$ unit.
The figure also shows the azimuthally averaged integrated intensity radial profiles of $^{12}$CO (red), $^{13}$CO (green), and C$^{18}$O (blue) line in the $\rm Jy~arcsec^{-2}~km~s^{-1}$ unit. We integrate the intensities of the CO isotopologue lines within the same velocity range from $\rm\sim -1~km~s^{-1}$ to $\rm\sim12~km~s^{-1}$. The color-shaded regions indicate the uncertainties of the (integrated) intensity profiles due to the azimuthal average.
As shown in Figure \ref{fig:figure1}(a), there is a dust gap at $\rm r\sim90$ au and a dust ring at $\rm r\sim120$ au. Compared with the rapid drop of the dust continuum intensity around 90 au, the CO isotopologue emission lines are extended to $\rm r\sim240$ au.

Figure \ref{fig:figure2}(b) presents the ratios between the azimuthally averaged integrated intensities of the CO isotopologue lines: $^{13}$CO/$^{12}$CO (green), C$^{18}$O/$^{12}$CO (red), and C$^{18}$O/$^{13}$CO (blue). The color-shaded regions indicate the uncertainties of the line ratios due to those of azimuthally averaged integrated CO intensities in Figure \ref{fig:figure2}(a). 
We note that even if we derive the line ratios by dividing the moment 0 maps first and then averaging azimuthally, the results are identical inside 200 au. In the region outside of 200 au, the signals are weak so that the line ratios have very large uncertainties. 

The line ratios show that the $^{12}$CO and $^{13}$CO lines are optically thick toward the CR Cha disk because the measured line ratios between three CO isotopologue emission lines are larger than 0.2 at all radii, which is larger than the typical abundance ratio of $^{13}$CO:$^{12}$CO = 1:67 and C$^{18}$O:$^{13}$CO = 1:7 in protoplanetary disks \citep[e.g.,][]{Qi2011}. When the molecular lines are optically thin, the line ratios should be almost the same as the abundance ratios because the line strength is proportional to the column density of the molecules. Meanwhile, when the lines are optically thick, the intensity at the line center becomes the same as the blackbody radiation and irrelevant to the molecular abundances. Thus, the line ratios do not need to be close to the abundance ratios.


\subsection{The peak brightness temperatures of the CO isotopologue lines}
\label{subsec:Tgas}

Since $^{12}$CO and $^{13}$CO lines are optically thick, we can adapt the brightness temperature of these lines as a gas temperature. To estimate the CO gas temperature, we make the moment 8 maps of $^{12}$CO and $^{13}$CO emission lines using \texttt{immoments} command in \texttt{CASA}. The moment 8 map shows the peak intensity (I$_{\rm \nu,peak}$) at each pixel in the data cube. 
For making the moment 8 maps, we used the data cube of $^{12}$CO and $^{13}$CO emission lines without dust continuum subtraction. The line plus continuum intensity should be used in order to derive the gas temperature from the peak brightness temperature of an optically thick line \citep[e.g.,][]{Weaver2018}.

Figure \ref{fig:figure4}(a) and (b) present the moment 8 maps of $^{12}$CO and $^{13}$CO line emissions including the dust continuum. The white and cyan ellipses are the same as those in Figure \ref{fig:figure1}. The white solid contour indicates the location of the bump at 130 au in the peak intensity and the peak brightness temperature of $^{13}$CO line emission in Figure \ref{fig:figure4}(c) and (d). 
We azimuthally average only the southern half-disk of the moment 8 maps to derive the I$_{\rm \nu,peak}$ radial profiles for avoiding the absorption feature. Figure \ref{fig:figure4}(c) shows the derived I$_{\rm \nu,peak}$ radial profiles of $^{12}$CO (red) and $^{13}$CO (blue). The color-shaded regions present the uncertainties due to the azimuthal average. The $\rm I_{B,peak,^{13}CO}$ profile shows a small bump around $\rm r\sim130$ au, corresponding to the radius of the dust ring beyond the dust gap. 

Using the I$_{\rm \nu,peak}$ profiles, we derive the peak brightness temperature (T$_{\rm B,peak}$) by assuming $\rm I_{ \nu,peak}=B_{\rm \nu}(T_{\rm B,peak})$ where B$_{\nu}$ is the Planck function at a given frequency $\nu$. The derived T$_{\rm B,peak}$ profiles are presented in Figure \ref{fig:figure4}(d). The color-shaded regions are the uncertainties of T$_{\rm B,peak}$ due to those of I$_{\rm \nu,peak}$ in Figure \ref{fig:figure4}(c). As the same as I$_{\rm ^{13}CO,peak}$ profile, the derived T$_{\rm B,peak}$ profile from $^{13}$CO line also shows a small bump at $\sim130$ au.

To quantify the temperature bump at $\sim130$ au, we do the least square fit to the derived T$_{\rm B,peak}$ profiles using the power-law profile with a single Gaussian,
\begin{equation}
 {\rm T_{\rm B,peak}}(r) = A \left(\frac{r}{\rm 50~au}\right)^{B} + C~{\rm exp}\left(-\frac{(r-D)^{2}}{2 E^{2}}\right)
\label{eq:eq21}
\end{equation}
where $A$, $B$, $C$, $D$, and $E$ are the fitting parameters. Using the data within $\rm 20~au\leq r\leq 200~au$, we obtain the fitting parameters, as summarized in Table \ref{tab:table3}. The fitting parameters show that the small bump of T$_{\rm B,peak,^{13}CO}$ profile at $\rm r\sim127$ au has $\sim 4.3$ K amplitude, corresponding to $\sim1.4\sigma$ at this radius ($1\sigma\approx3$ K). The dashed lines in Figure \ref{fig:figure4}(d) present the fitting results of T$_{\rm B,peak}$ profiles. 

The T$_{\rm B,peak}$ profile derived from I$_{\rm ^{12}CO,peak}$ also has a small bump at $\rm r\sim82$ au, the inner edge of the dust gap. It may be related to the slope change of the dust continuum intensity. 
The location of the photosphere of the $^{12}$CO line could change inside and outside of the dust rich disk. The disk scale height could also change there (see Section \ref{subsec:gap origin}).
The difference between the $^{12}$CO and $^{13}$CO brightness temperatures can be a hint for the temperature and density distribution of the gas disk. We will discuss it more in Section \ref{subsubsec:dust trap model}.

\section{Discussion}
\label{sec:discussion}

\subsection{The dust gap in the CR Cha disk}
\label{subsec:SD derivation}

To analyze the dust gap structure of the CR Cha disk, we derive the  radial profile of dust surface density ($\Sigma_{\rm dust}$) from the observed dust continuum intensity. 
We derive the optical depth at 225 GHz, $\tau_{\rm 225GHz}$, using the radiative transfer equation, $\rm I_{\rm cont}=B_{\rm 225GHz}(T_{\rm dust})\left(1-exp(-\tau_{\rm 225GHz})\right)$, where we simply assume that the dust temperature is equal to the gas temperature derived from the moment 8 maps of the $^{12}$CO or $^{13}$CO line in Figure \ref{fig:figure4}(d), that is, $\rm T_{\rm dust}=T_{\rm gas}=T_{\rm B,peak}$. We note that since both CO lines are optically thick, we can use the derived T$_{\rm B,peak}$ profiles as the gas temperature (T$_{\rm gas}$) profiles of the $^{12}$CO and $^{13}$CO line emitting regions. And then we derive the dust surface density $\Sigma_{\rm dust}$ using the equation $\tau_{\rm 225GHz}=\kappa_{\rm 225GHz}\Sigma_{\rm dust}$ where $\rm \kappa_{\rm 225GHz}=2.3~cm^{2}~g^{-1}$ \citep[e.g.,][]{Beckwith1990} is the dust opacity at 225 GHz. 

\begin{figure}[t]
\centering
\includegraphics[width=0.45\textwidth]{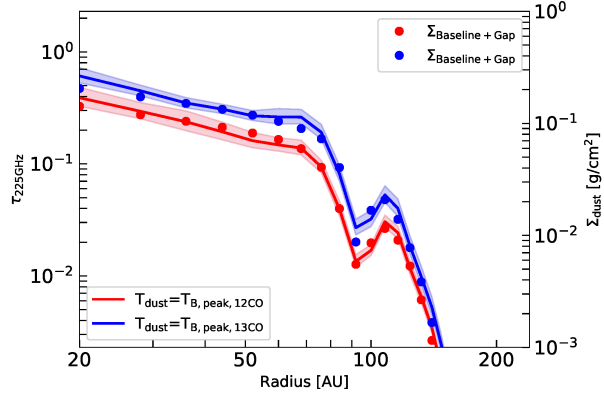}
\caption{The radial profiles of optical depth at 225 GHz ($\tau_{\rm 225GHz}$) and the dust surface density ($\Sigma_{\rm dust}$), where  $\tau_{\rm 225GHz}=\kappa_{\rm 225GHz}\Sigma_{\rm dust}$ with the constant $\kappa_{\rm 225GHz}$=2.3 cm$^{2}$ g$^{-1}$. The solid lines present the derived $\tau_{\rm 225GHz}$ and $\Sigma_{\rm dust}$ profiles when we adopt the peak brightness temperature of $^{12}$CO (red) and $^{13}$CO (blue) as dust temperature. The color-shaded regions indicate the unceratinties propagated from those of the peak brightness temperature in Figure \ref{fig:figure4}(d). The Gaussian fitting results of the baseline plus dust gap expressed in Equation (\ref{eq:eq1}) and (\ref{eq:gap}) are shown as dotted lines with the same color indications. The fitting parameters summarized in Table \ref{tab:table4} show that the dust gap is centered at $\rm r\sim90$ au with a width of $\sim8$ au on average. The depth of the gap at the center is $\sim20\%$ on average.}
\label{fig:figure5}
\end{figure}

The solid lines in Figure \ref{fig:figure5} present the derived $\tau_{\rm 225GHz}$ and $\Sigma_{\rm dust}$ profiles when the T$_{\rm dust}$ equals T$_{\rm B,peak,^{12}CO}$ (red) and T$_{\rm B,peak,^{13}CO}$ (blue). The color-shaded regions indicate the uncertainties of $\tau_{\rm 225GHz}$ and $\Sigma_{\rm dust}$ propagated from those of the peak brightness temperature in Figure \ref{fig:figure4}(d). 
We note that although we simply assume $\rm T_{dust}=T_{gas}$, the dust temperature near the disk midplane could be lower than the gas temperature of the $^{13}$CO or $^{12}$CO line emitting regions (see Section \ref{subsubsec:dust trap model}). Thus, the derived dust surface density could be lower than the actual dust surface density.

We adopt a single Gaussian function to estimate the dust gap structure at $\rm r\sim90$ au. As the baseline of dust surface density profile, we adopt the power-law function with an exponential tail,  
\begin{equation}
\Sigma_{\rm d, baseline} = A \left(\rm \frac{r}{50~au}\right)^{-0.5}~{\rm exp}\left[\rm - \left(\frac{r}{100~au}\right)^{4} \right]
\label{eq:eq1}
\end{equation}
where $A=\rm 0.09~g~cm^{-2}$ for $\rm T_{\rm dust}=T_{\rm B,peak,^{12}CO}$ and $A=\rm 0.13~g~cm^{-2}$ for $\rm T_{\rm dust}=T_{\rm B,peak,^{13}CO}$. We note that the terms of Equation (\ref{eq:eq1}) is derived by the least square fitting to reproduce the shape of the profiles. And then, we choose concrete numbers within the uncertainties of the fitting parameters. After subtracting the baseline from the derived $\Sigma_{\rm dust}$ profiles, the gap structure is fitted using a single Gaussian function,
\begin{equation}
\rm \Sigma_{\rm d, gap} = \Sigma_{peak} \exp\left[ \frac{(r-r_{cent})^{2}}{2(\Delta r)^{2}}\right].
\label{eq:gap}
\end{equation} 
Table \ref{tab:table4} summarizes the Gaussian fitting parameters for the gap structure. The uncertainties of the fitting parameters are more than 500\% for all parameters due to the small number of data points in the gap: only $2-3$ data points are included in the gap width $\sim8$ au due to 8 au width of annuli for the azimuthal average under the beam size of $\sim0.087''$ ($\sim16$ au at $d\sim187.5$ pc).

The dotted lines in Figure \ref{fig:figure5} presents the fitting results of the baseline plus dust gap. 
For both cases, the fitting results show that the dust gap is centered at $\rm r\sim90$ au and has the width of  $\sim8$ au on average. The depth of the gap is measured as $\left(\Sigma_{\rm d, baseline}(\rm r_{cent})-\Sigma_{\rm peak}\right)/\Sigma_{\rm d, baseline}(\rm r_{cent})$, the ratio between the bottom of the gap and the baseline value at the gap center $\rm r_{cent}$. The derived gap depth is $\sim20\%$ on average. We note that the derived gap depth should be regarded as the upper limit due to the larger synthesized beam size ($\sim16$ au) than the estimated gap width $\sim8$ au. The observed intensity profile can be reproduced by a deeper and narrower gap than those derived from our observation. As an example, the observation of HD 163296 disk with high angular resolution \citep[$\sim0.05''$,][]{Isella2018} shows that the gap is deeper than that observed with low angular resolution \citep[$\sim0.2''$,][]{Isella2016PhRvL}.

\begin{figure*}[t]
\centering  
\subfiguretopcaptrue
\subfigure[$\tau_{\rm C^{18}O,peak}$ radial profile]{\label{fig:6a}\includegraphics[width=0.45\textwidth]{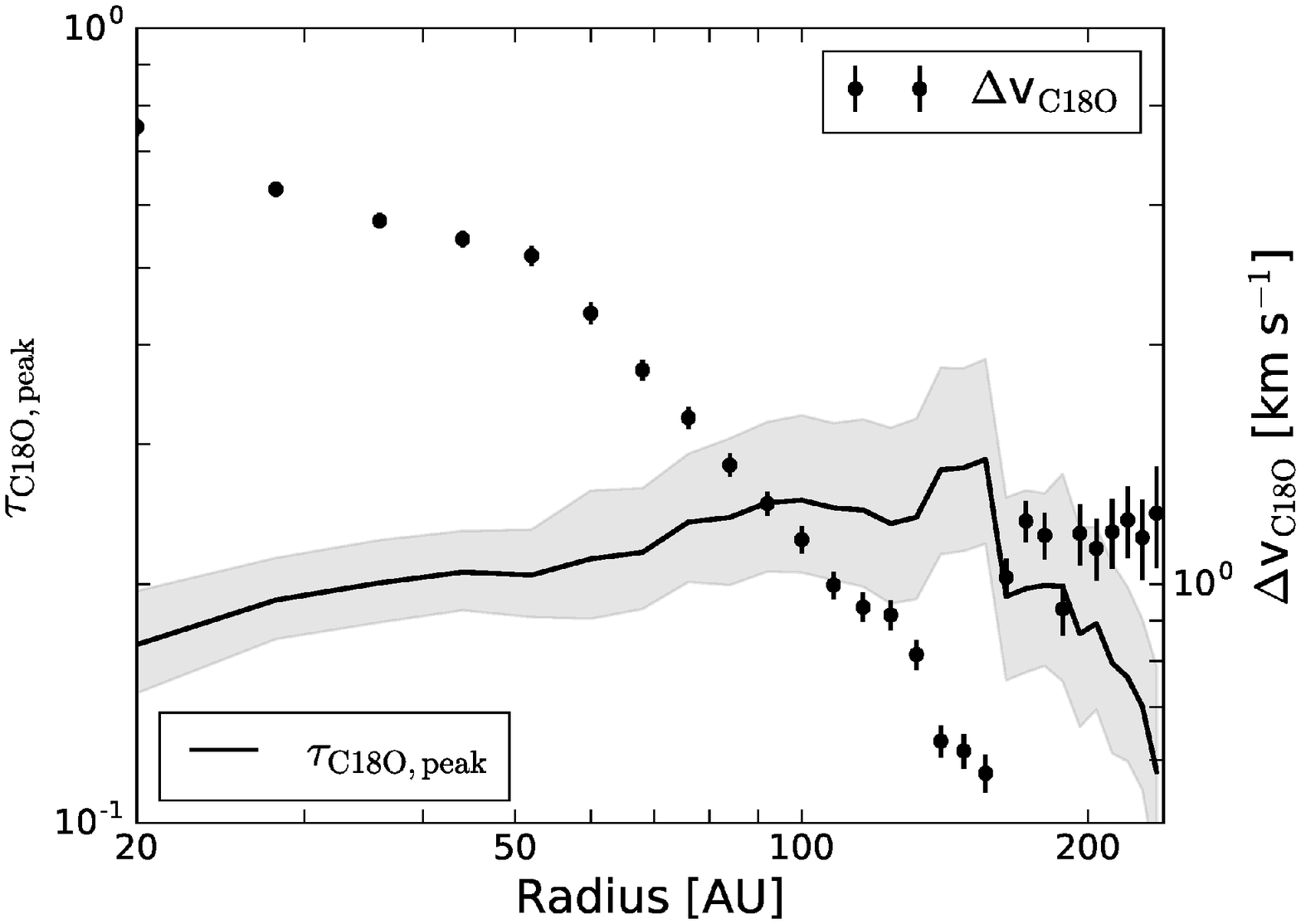}} \hfill
\subfigure[N$_{\rm C^{18}O}$ radial profile]{\label{fig:6b}\includegraphics[width=0.48\textwidth]{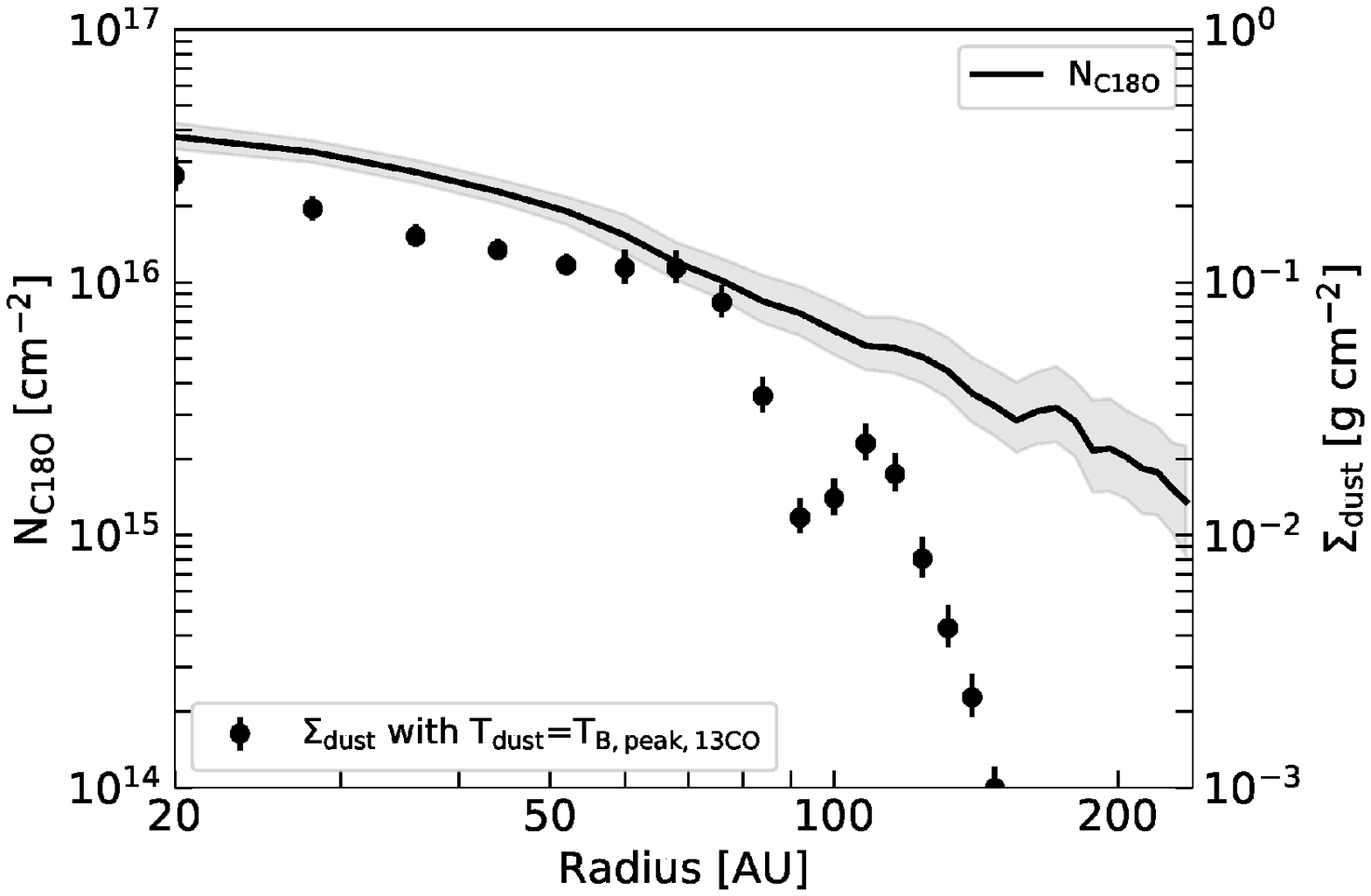}}
\hfill
\subfiguretopcapfalse
\caption{(a) The radial profile of $\tau_{\rm C^{18}O,peak}$ (solid line) with the uncertainty (gray-shaded regions). The dotted line with error bars indicates the fitted width of the averaged C$^{18}$O line spectra at each annulus. The optical depth in the inner disk decreases due to the line broadening caused by the high rotation velocity in the very inner disk. This line broadening leads to small Gaussian peak. 
(b) The radial profiles of the C$^{18}$O column density N$_{\rm C^{18}O}$ (solid line) with the uncertainty (gray-shaded region). The dotted line with error bars indicates the dust surface density profile derived using $\rm T_{\rm dust}=T_{\rm B,peak,^{13}CO}$ in Section \ref{subsec:SD derivation}. The gas disk is extended much further than the dust disk.}
\label{fig:figure6}
\end{figure*}

\begin{table}[t]
\begin{center}
\caption{The Gaussian fitting parameters for the dust gap}
\begin{tabular}{c|c|c|c|c}
\hline \hline \noalign {\smallskip}
T$_{\rm dust}$ & r$_{\rm cent}$ & $\Delta$r & $\Sigma_{\rm peak}$ & Depth \\ 
 & [au] & [au] & [g cm$^{-2}$] & [\%] \\ 
\hline \noalign {\smallskip}
T$_{\rm B,peak.^{12}CO}$ & 88.94 & 9.74 & 0.028 & $\sim21.71$ \\ 
T$_{\rm B,peak.^{13}CO}$ & 91.47 & 6.72 & 0.038 & $\sim19.94$ \\ 
\hline \noalign {\smallskip}
\end{tabular}
\end{center}
\label{tab:table4}
\end{table}

\subsection{ {\rm C$^{18}$O} gas column density}
\label{subsec:column_density}

In this section, we analyze the CO gas radial distribution in the CR Cha disk. We adopt the following equation in \cite{Yamamoto2017} to derive the C$^{18}$O column density (N$_{\rm C^{18}O}$):
\begin{equation}
\begin{split}
\int{ \tau_{\nu} \rm dv} & \approx \sqrt{2\pi}~\tau_{\rm \nu,peak}\Delta \rm v \\ 
= & \frac{8\pi^{3}S\mu_{0}^{2}}{3h U(\rm T_{ gas})} \left[ \exp \left(\frac{h\nu}{ k_{ B}\rm T_{gas}}\right) -1\right] \\
& \times \exp\left(-\frac{E_{u}}{k_{B}\rm T_{gas}} \right) N_{\rm molecule} \\
\end{split}
\label{eq:eq2}
\end{equation}
where $\tau_{\rm \nu,peak}$ is the peak optical depth of the molecular line at the line center, $\Delta$v is the width of the molecular line, $S$ is the molecular line strength, $\mu_{0}$ is the permanent dipole moment of the molecule, $h$ is the Planck constant, $U(\rm T_{\rm gas})$ is the rotational partition function of the molecular species at a given gas temperature T$_{\rm gas}$, $\nu$ is the rest frequency of the molecular transition line, $k_{\rm B}$ is the Boltzmann constant, $E_{u}$ is the upper state energy level, and $N_{\rm molecule}$ is the column density of the molecule, respectively. We simply assume that the line optical depth has a Gaussian profile with the peak value of $\tau_{\rm \nu,peak}$ and the width of $\rm \Delta v$ for $\rm \int \tau_{\nu} dv\approx\sqrt{2\pi}\tau_{\rm \nu,peak}\Delta v$. 

For C$^{18}$O $J=2-1$ transition line, we adopt $\rm \nu=219.5603~GHz$, $S\mu_{0}^{2}=0.02440~{\rm D}^{2}\approx2.440\times10^{-51}~{\rm J\cdot m^{3}}$, and $E_{ u}/k_{B}=15.80580~\rm K$ from the Leiden Atomic and Molecular Database \citep[][]{Schoier2005} and Cologne Database for Molecular Spectroscopy \citep{Muller2001CDMS}. The rotational partition function $U(\rm T_{\rm gas})$ of C$^{18}$O molecule is approximated as $U(\rm T_{\rm gas})\approx0.38\left(\rm T_{\rm gas}+0.88\right)$ \citep{Mangum2017}. 

As a first step, we derive the peak optical depth at the C$^{18}$O line center ($\tau_{\rm C^{18}O,peak}$) using the radiative transfer equation, $\rm I_{\rm C^{18}O,peak}=B_{\rm C^{18}O}(T_{\rm gas}) \left[1-\exp(-\tau_{\rm C^{18}O,peak})\right]$ with $\rm T_{gas}=T_{B,peak,^{13}CO}$, the blue line in Figure \ref{fig:figure4}(d). The I$_{\rm C^{18}O,peak}$ is derived from the Gaussian fitting of the azimuthally averaged C$^{18}$O line spectra of every 8 au width annulus within $\rm r\lesssim240~au$. 
When we azimuthally average the C$^{18}$O line spectra, we shift the velocity offset of the line at each point in the annulus (mainly due to Doppler shift by the Keplerian rotation) by using the moment 1 map of the C$^{18}$O line. 

From the azimuthally averaged line spectra, we obtain the peak C$^{18}$O line intensity (I$_{\rm C^{18}O,peak}$) and the width ($\Delta$v$_{\rm C^{18}O}$) by Gaussian fitting. 
The peak optical depth of C$^{18}$O line ($\tau_{\rm C^{18}O, peak}$) is presented in Figure \ref{fig:figure6}(a) as solid line. The gray-shaded region shows the uncertainty of the $\tau_{\rm C^{18}O, peak}$. The dotted line with error bars in the same figure shows the fitted line width $\Delta$v$_{\rm C^{18}O}$. 
The decrease of $\tau_{\rm C^{18}O}$ in the inner disk is caused by the line broadening. Since the Keplerian rotation velocity is higher and the gradient of velocity along the line of sight is steeper, the beam averaged line width becomes broader in the inner disk. Due to this line broadening, the $\tau_{\rm C^{18}O, peak}$ profile is nearly flat up to $\rm r\sim160$ au but the line width rapidly decreases. In the outer disk of $\rm r\gtrsim160~au$, the Gaussian fitting is affected by the noise next to the line signal making broader line width and smaller peak intensity. It makes the drop of $\tau_{\rm C^{18}O,peak}$ at $\rm r>160~au$. 

Finally, we derive the C$^{18}$O column density (N$_{\rm C^{18}O}$) by assigning the derived $\tau_{\rm C^{18}O,peak}$, $\rm \Delta v_{C^{18}O}$, and $\rm T_{gas}=T_{B,peak,^{13}CO}$ into Equation (\ref{eq:eq2}). 
Figure \ref{fig:figure6}(b) shows the derived N$_{\rm C^{18}O}$ profile as a solid line and their uncertainty as a gray-shaded region. Compared with the derived dust surface density profile, the dotted line with error bars, the C$^{18}$O gas disk is much extended than the dust disk. There is no gas gap around the dust gap at $\rm r\sim90$ au due to the large synthesized beam size of the line observation. 



\begin{figure*}[t]
\centering  
\subfiguretopcaptrue
\subfigure[The normalized dust-to-CO-gas mass ratio]{\label{fig:7a}\includegraphics[width=0.45\textwidth]{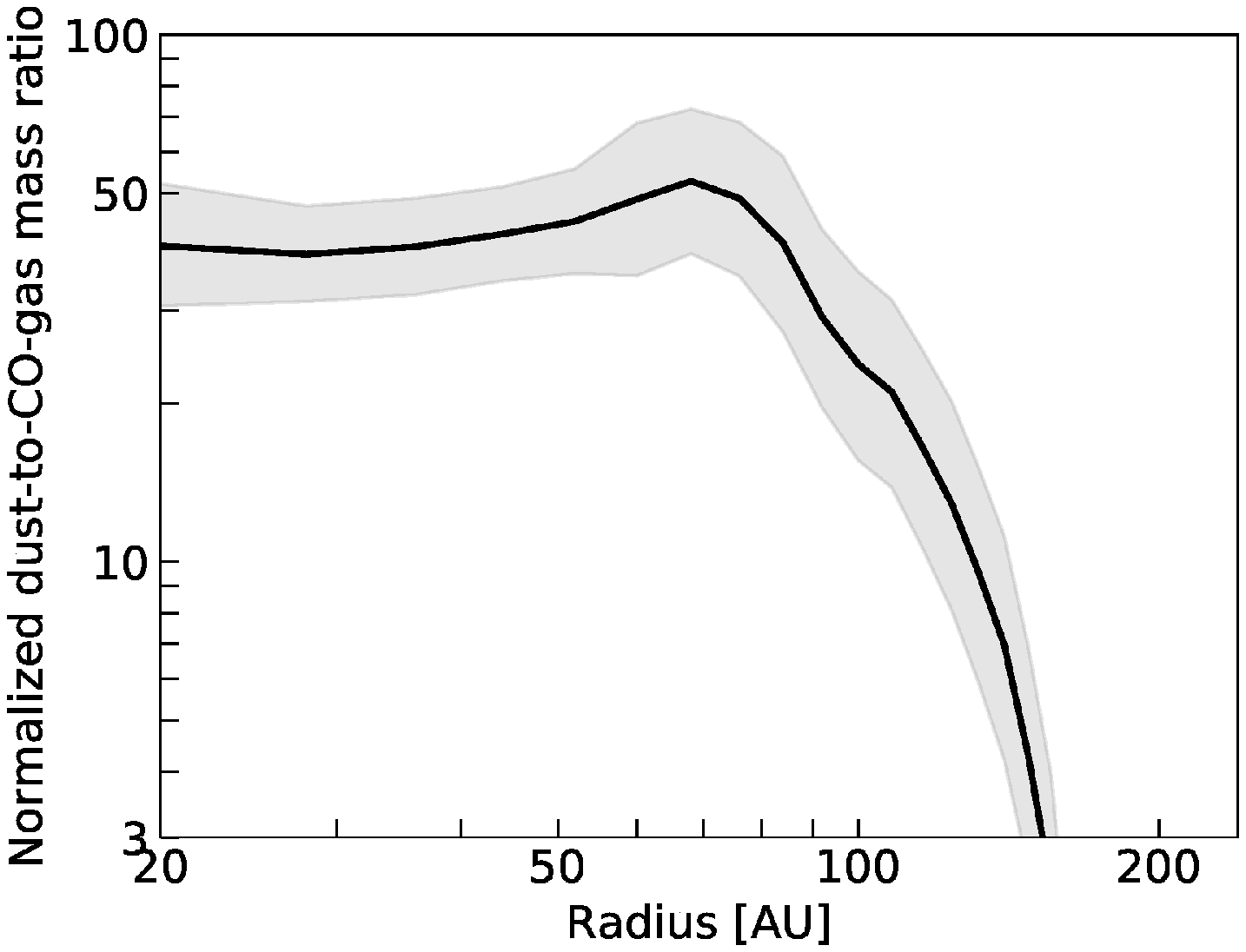}} \hfill
\subfigure[Toomre Q parameter]{\label{fig:7b}\includegraphics[width=0.45\textwidth]{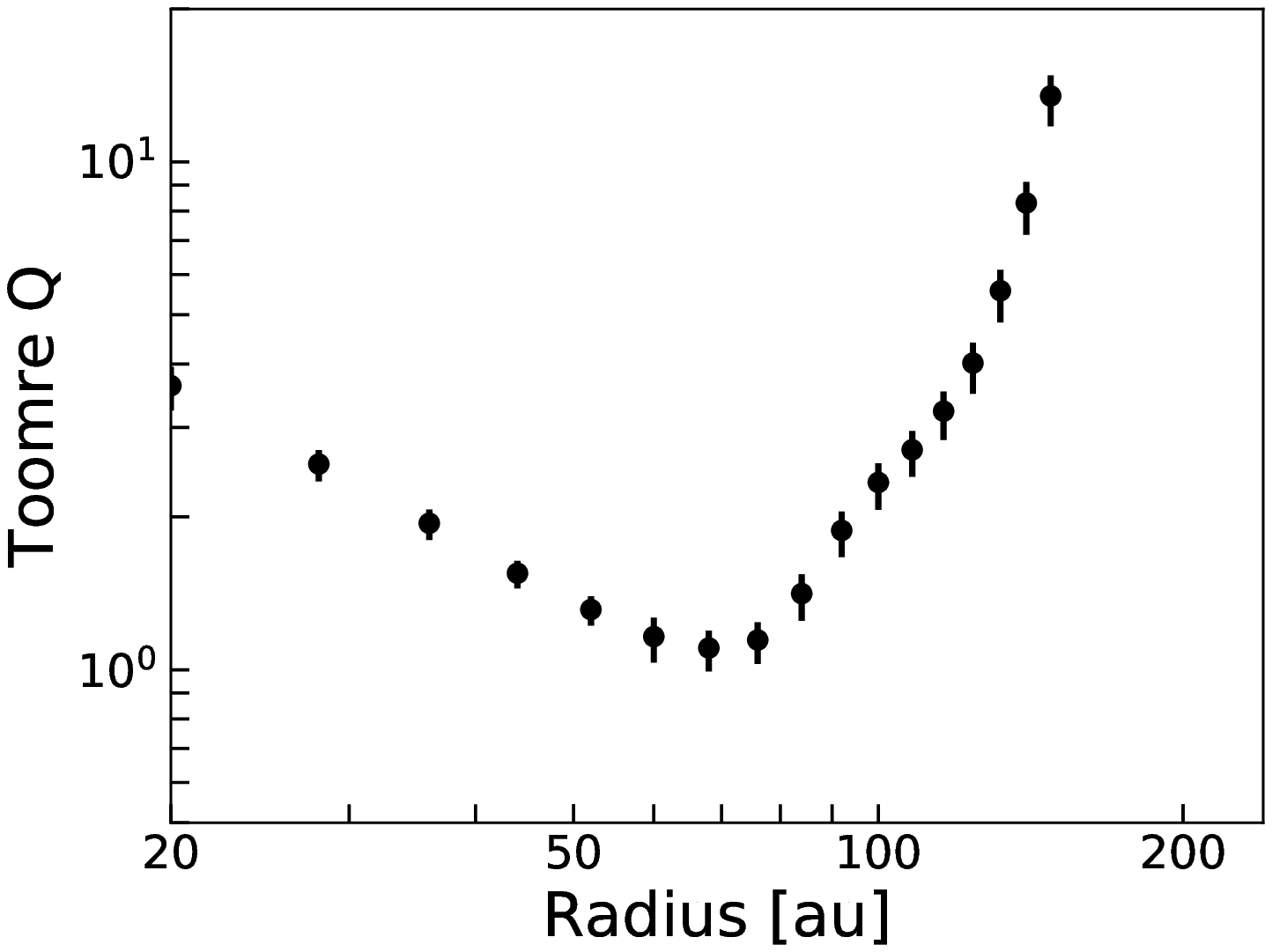}}
\hfill
\subfiguretopcapfalse
\caption{(a) The radial profile of dust-to-CO-gas mass ratio ($\epsilon/X$) normalized by the typical ratio in the interstellar medium ($\epsilon_{0}/X_{0}$). The $\Sigma_{\rm dust}$ profile is convolved with the same beam size of C$^{18}$O line image for deriving the dust-to-CO-gas mass ratio. The gray-shaded region indicates the uncertainty of the ratio. The disk region in $\rm r\lesssim80~au$ has $\sim50$ times higher dust-to-CO-gas mass ratio than the typical interstellar medium. Outside of the dust ring at 120 au, the ratio rapidly decreases toward the unity. 
(b) The Toomre's Q parameter of the disk derived from the $\Sigma_{\rm dust}$ in Figure \ref{fig:figure6}(b) and the assumption of $\Sigma_{\rm disk}\approx\Sigma_{\rm gas}=100\Sigma_{\rm dust}$. The region $\rm r\lesssim80~au$ has $\rm Q\approx1-3$, which is marginally gravitationally unstable disk.}
\label{fig:figure9}
\end{figure*}

\subsection{The dust-to-CO-gas mass ratio}
\label{subsec:dustgas}

The dust-to-gas mass ratio is one of the important parameters to understand the physical and chemical conditions of the protoplanetary disks. Based on the radial profiles of dust surface density and C$^{18}$O column density in Section \ref{subsec:SD derivation} and \ref{subsec:column_density}, we derive the dust-to-CO-gas mass ratio of the CR Cha disk. 

When we adopt the factor $X\equiv\rm N_{CO}/N_{H_{2}}$, the ratio of the number density between CO and H$_{2}$ molecules, and $\epsilon\equiv\Sigma_{\rm dust}/\Sigma_{\rm H_{2}}$, the dust to gas mass ratio, we derive the dust-to-CO-gas mass ratio
\begin{equation}
\frac{\epsilon}{X} = \frac{ \Sigma_{\rm dust} }{ \Sigma_{\rm H_{2}} } \frac{\rm N_{\rm H_{2}} }{ \rm N_{\rm CO} } 
= \frac{ \Sigma_{\rm dust} }{\rm 2m_{H} N_{CO} }
= 14~\frac{ \Sigma_{\rm dust} }{ \Sigma_{\rm CO} }
\end{equation}
where $\rm \Sigma_{H_{2}}=2m_{H}N_{H_{2}}$, $\rm \Sigma_{CO}=28m_{H}N_{CO}$, and $\rm m_{\rm H}=1.67\times10^{-24}~g$ is the mass of hydrogen atom. If we adopt the typical abundance ratio of $\rm ^{12}CO/C^{18}O$ $=444$ \citep[e.g.,][]{Qi2011}, the CO number density is derived by the conversion of $\rm N_{CO}=444~N_{C^{18}O}$. We adopt the typical values in the ISM of $X_{0}=10^{-4}$ and $\epsilon_{0}=10^{-2}$ for the comparison to the typical dust-to-CO-gas mass ratio in the interstellar medium (ISM). 

Figure \ref{fig:figure9}(a) presents the radial profile of the normalized dust-to-CO-gas mass ratio $\left(\epsilon/X\right) / \left(\epsilon_{0}/X_{0}\right) $, indicating how the dust-to-CO-gas mass ratio of the CR Cha disk deviates from the typical ISM value. The gray-shaded region indicates the uncertainty of the dust-to-CO-gas mass ratio based on the uncertainties of the derived $\Sigma_{\rm dust}$ and N$_{\rm C^{18}O}$. We note that the $\Sigma_{\rm dust}$ profile is convolved with the same beam size of C$^{18}$O line image for deriving the dust-to-CO-gas mass ratio. 
In the disk radius of $\rm r\lesssim80~au$, the dust-to-CO-gas mass ratio is $\sim50$ times higher than the typical ISM value. 


The dust concentration by the radial drift of dust grains is one of the possible mechanisms to explain this high dust-to-CO-gas mass ratio in the inner disk. Here, we simply assume that the initial dust disk is extended to $\rm r=240$ au with a uniform dust surface density and the dust mass is conserved during the radial contraction.
If only dust grains drift inward down to $\rm r\sim90$ au from $\rm r=240$ au, $\sim9$ times dust concentration is acceptable. However, it is still not enough to reach $\sim50$ times higher dust-to-CO-gas mass ratio so that they may need other causes to increase the dust-to-CO-gas mass ratio. The CO gas depletion ($X=\rm N_{CO}/N_{H_{2}}<10^{-4}$) and/or gas dispersal ($\epsilon=\rm \Sigma_{dust}/\Sigma_{gas}>0.01$) may have to be considered as another possibility to explain this high dust-to-CO-gas mass ratio. 

We note that there are some uncertainties in the derived dust-to-CO-gas mass ratio. As mentioned in Section \ref{subsec:column_density}, the dust surface density could be underestimated because the temperature at the disk midplane, where optically thin dust continuum will be mainly emitted, is possibly lower than the temperature of the $^{13}$CO line emitting region in the disk surface. In that case, the dust-to-CO-gas mass ratio could be underestimated. Also, the dust opacity has large uncertainty, which, too, leads to uncertainty in derived dust surface density. Meanwhile, if the C$^{18}$O line is optically thick, the derived C$^{18}$O column density will be underestimated. Thus, the derived dust-to-CO-gas mass ratio could be overestimated.

Although we can reproduce the observed high dust-to-CO-gas ratio with dust concentration, gas dispersal, or CO depletion, this high dust-to-CO-gas mass ratio can trigger the gravitational or streaming instabilities. The disk stability by self-gravity is presented by the Toomre Q parameter \citep{Toomre1964},
\begin{equation}
Q\equiv\frac{c_{\rm s} \Omega_{\rm K}}{\pi G \Sigma_{\rm disk}},
\end{equation}
where $c_{\rm s}$ sound speed, $\Omega_{\rm K}$ orbital frequency, G gravitational constant, and $\Sigma_{\rm disk}$ disk surface density. The assumption of $\rm T=T_{\rm B,peak,^{13}CO}$ and $M_{\star}=1.657~M_{\odot}$ are used for deriving $c_{\rm s}=\sqrt{\rm kT/\mu m_{\rm H}}$ and $\Omega_{\rm K}=\sqrt{GM_{\star}/r^{3}}$. If the observed high dust-to-CO-gas mass ratio is explained only by the CO gas depletion and there is no gas dispersal or dust concentration ($\rm \Sigma_{disk}\approx\Sigma_{gas}=100\Sigma_{dust}$), the derived Toomre Q parameter is $\rm Q\sim1-3$ inside of $\rm r\sim100$ au, shown in Figure \ref{fig:figure9}(b), indicating that the disk is marginally unstable by self-gravity.
Meanwhile, if there is gas dispersal and/or dust concentration, Figure 9 in \cite{YangCC2017} shows that $\Sigma_{\rm dust}/\Sigma_{\rm gas}\gtrsim0.2$ triggers the streaming instability regardless of the grain sizes. It indicates that $(\epsilon/X)/(\epsilon_{0}/X_{0})\gtrsim20$ can trigger the streaming instability if we assume no CO gas depletion (that is, $X=N_{\rm CO}/N_{\rm H_{2}}=10^{-4}$). 
If these instabilities are triggered, the dust disk is rapidly dissipated by creating planetesimals or disk substructures like spiral arms and cannot stand for a long time \citep[e.g.,][]{Dong2015SI-spiral,Li2019}. Since it is not consistent with the observation, some assumptions used in the derivations of dust and gas surface density from our observations may not be realistic.


\subsection{The formation of dust gap-ring structure}
\label{subsec:gap origin}

In this section, we discuss the possible formation mechanisms of the dust gap-ring structure in the outer region of the dust disk. Here, we introduce two possible mechanisms: gap opening by a planet and dust concentration on the gas pressure bump at the outer edge of the dust disk.

\subsubsection{Gap opening by a planet}

The surface density distribution of the gap formed by a planet is well studied by numerical simulations in previous studies. Here, we compare the surface density distribution around the gap obtained from the observations with the numerical models to constrain the planet mass which can reproduce the observed gap structure.

We adopt the model of \cite{Rosotti2016MNRAS_459_2790R} to estimate the minimum planet mass for reproducing the dust gap structure in the CR Cha disk. They suggest that the minimum planet mass can perturb the dust surface density is 
\begin{equation}
M_{\rm p,min}\sim15~M_{\oplus} \left(\frac{H/R}{0.05}\right)^{3} \left(\frac{M_{\ast}}{1~M_{\odot}}\right)
\label{eq:mass_min}
\end{equation}
where $H$ is the disk vertical scale height, $R$ is the radial location of the planet, and $M_{\ast}$ is the central star mass. To evaluate $H/R$, we use $\rm T=22$ K at the gap center of $\rm r_{cent}=R=91.47$ au, which are obtained from $^{13}$CO line emission and from the dust surface density distribution. Assigning $M_{\ast}=1.657~M_{\odot}$ \cite[]{Villebrun2019} and $H/R=0.0683$ into Equation (\ref{eq:mass_min}), we obtain $M_{\rm p,min}\sim0.2~M_{\rm J}$. This is the lower limit of the planet mass for creating a dust gap in the CR Cha disk. 

Next, we evaluate the planet mass in the gap through the fitting of the observed gap structure using Equation (6)-(10) in \cite{Kanagawa2017}. 
In \cite{Kanagawa2017}, the surface density profile is controlled by two parameters, $K$ and $K'$:
\begin{equation}
 K = \left( \frac{M_{\rm p}}{M_{\ast}} \right)^2
\left( \frac{h_{\rm p}}{R_{\rm p}} \right)^{-5}
\alpha^{-1},
\label{eq:eq7}
\end{equation} 
\begin{equation}
 K' = K \left(\frac{h_{\rm p}}{R_{\rm p}}\right)^2,
\end{equation}
where $M_{\rm p}$ is the planet mass, $M_{\ast}$ is the central star mass, $h_{\rm p}$ is the disk scale height at the planet orbital radius $R_{\rm p}$, and $\alpha$ is the dimensionless parameter for the strength of the turbulence \cite[]{SS1973}. To evaluate these parameters, we use the same parameters used above: $\rm T=22~K,~r_{cent}=91.47~au$. The gap center is also used as the planet orbital radius $R_{\rm p}$. 
Using these values, we find that the parameters $K$ and $K'$ depend on a single parameter $\alpha^{-1}(M_{\rm p}/M_{\ast})^2$.

\begin{figure}[t]
\centering
\includegraphics[width=0.45\textwidth]{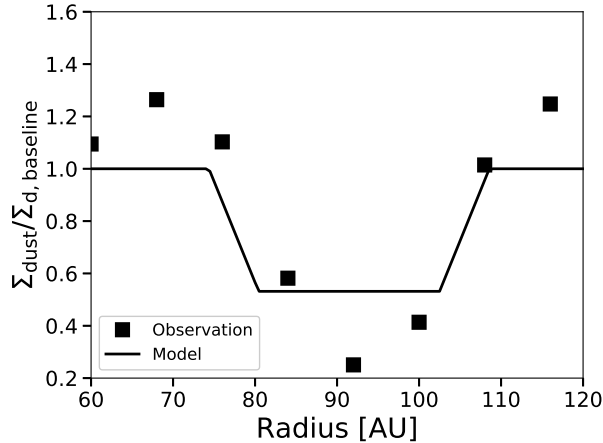}
\caption{Comparison between the surface density profiles derived from our observation (filled squares) and predicted by the gap opening model (solid line). The vertical axis is the surface density normalized by the baseline function expressed as Equation \ref{eq:eq1}. 
}
\label{fig:gap}
\end{figure}

Figure \ref{fig:gap} shows the comparison between the surface density distributions derived from the observation (filled square) and predicted by the model of gap opening by a planet (solid line). We note that we normalize the observed dust surface density by the baseline function Equation (\ref{eq:eq1}) for the fitting to the model. 
The best fit value is $\alpha^{-1}(M_{\rm p}/M_{\ast})^2 = 3.3\times10^{-5}$. Thus, the planet mass can open the observed gap structure is
\begin{equation}
 M_{\rm p} = 0.99 M_{\rm J}\left(\frac{\alpha}{0.01}\right)^{0.5},
\end{equation}
where $M_{\rm J}$ is the Jupiter mass. Note that the planet mass estimated here has a large uncertainty caused by the resolution of the observations and the dust evacuation from the gap as mentioned above. 

We note that there are two uncertainties for this planet mass estimation. First, the observed gap structure is not well resolved by the beam as mentioned in Section \ref{subsec:SD derivation}. Thus, the gap in the dust surface density would be deeper than that obtained from the observations. Second, the depth of dust gap may be much deeper than that of the gas gap. We assume that the dust gap structure obtained from our observation can trace the gas gap structure. This assumption will be justified when the dust grain size is small enough so that the dust grains are tightly coupled with the gas. 
In contrast, dust particles will be evacuated from the gas gap if the dust particles grow to about mm size \cite[]{Juan_Ovelar2013,Dong2015gap}. In this case, we obtain the depth of the gas gap larger than the real, which concludes a more massive planet mass to reproduce the observed gap depth.

We also estimate the upper limit of the planet mass from the cases that the gas gap is too shallow or too wide to be non-detection within the uncertainties of our observation. The ratio of the derived lower and upper limit of N$_{\rm C^{18}O}$ at 100 au is $\sim0.6$. If we assume that the upper limit is the baseline value and the lower limit is the bottom of the gas gap, 
\begin{equation}
\frac{\Sigma_{\rm lower}}{\Sigma_{\rm upper}} = \frac{1}{1+0.04 K} > 0.6.
\label{eq:gasdepth}
\end{equation}
When we assign the $R_{\rm p}=91.47$ au, $\rm T=22$ K, and $M_{\ast}=1.657~\rm M_{\odot}$ into Equation (\ref{eq:gasdepth}), the estimated upper limit of the planet mass is 
\begin{equation}
 M_{\rm p} < 0.87M_{\rm J} \left(\frac{\alpha}{0.01}\right)^{1/2}.
\end{equation}
Similarly, if the gas gap has a comparable width with the synthesized beam size of our observation, we also cannot detect the gas gap in the N$_{\rm C^{18}O}$ profiles. The relation between the gap width and the planet mass is expressed as 
\begin{equation}
\Delta R = 0.33 K'^{1/4}R_{\rm p}.
\end{equation}
Considering the beam size of $\sim0.2"$ corresponding to $\sim37.6$ au at 187.5 pc in physical scale, the upper limit of the planet mass is derived as
\begin{equation}
M_{\rm p} < 1.2 M_{\rm J} \left(\frac{\alpha}{0.01}\right)^{1/2}.
\end{equation}

Those values conclude that about Jupiter-mass planets are required to reproduce the observed dust gap structures regardless of the existence of the gas gap at the dust gap location.
The scattered light imaging is a good next step to investigate the planet mass estimation and gap structures by their higher spatial resolutions and direct tracing of small dust grains that are generally well-couple with the gas \citep[e.g.,][]{Dong2017ApJ_835_146D}.

The temperature bump around 130 au could be related to the heating of the outer wall of the gap. According to the previous studies, the gap opening by a planet changes the temperature distribution around the gap structure \cite[e.g.,][]{Jang-Condell2012, Turner2012}. Due to the smaller surface density inside the gap structure, this region can be heated by the stellar radiation which can penetrate to a deeper layer of the disk.

\begin{figure*}[t]
\centering  
\subfiguretopcaptrue
\subfigure[The $\Sigma_{\rm dust}$ and N$_{\rm C^{18}O}$ profiles]{\label{fig:sigma_d}\includegraphics[width=0.45\textwidth]{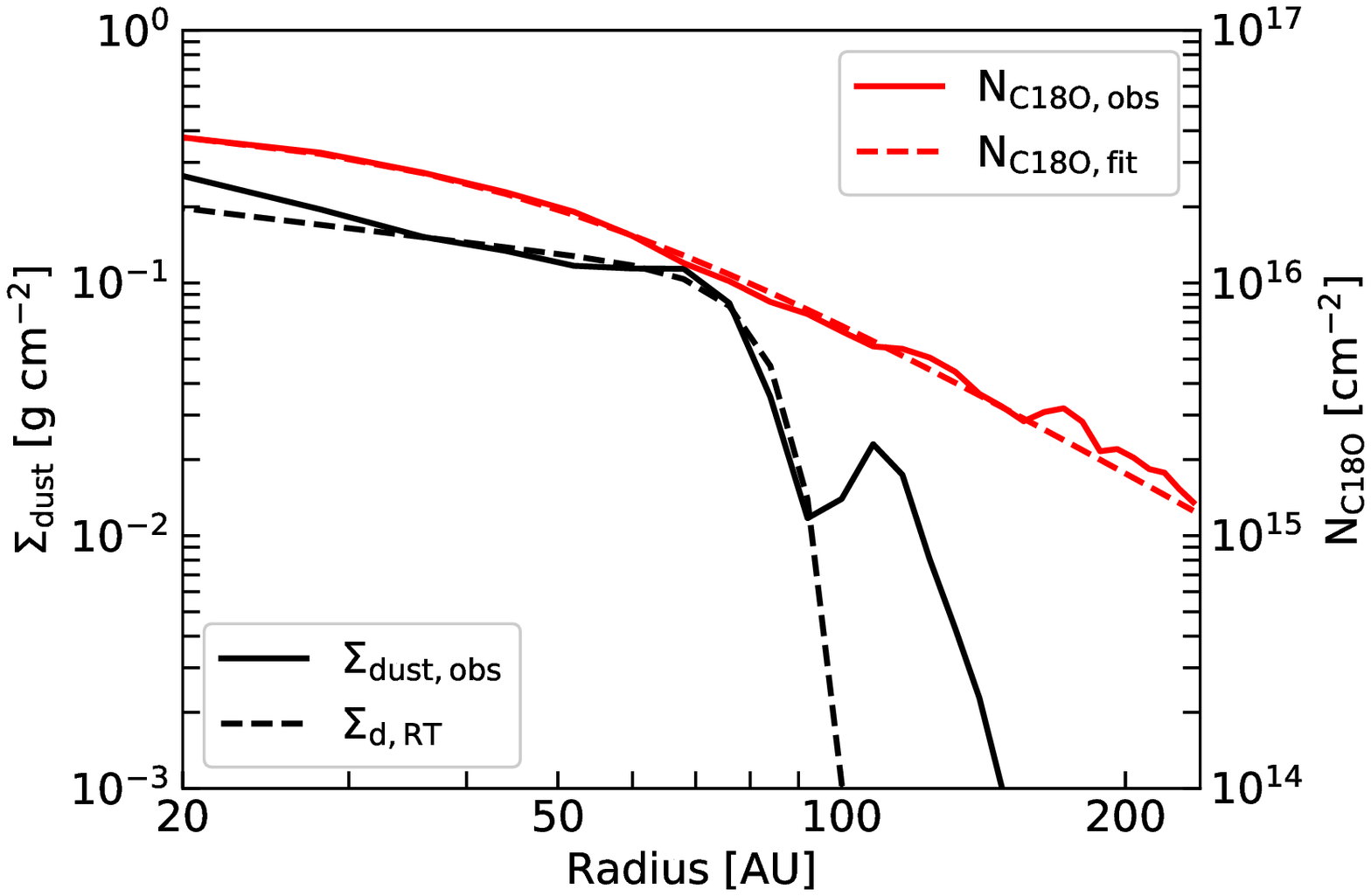}}
\subfigure[The T$_{\rm gas}$ distribution in the disk r-z plane]{\label{fig:temp_rz}\includegraphics[width=0.45\textwidth]{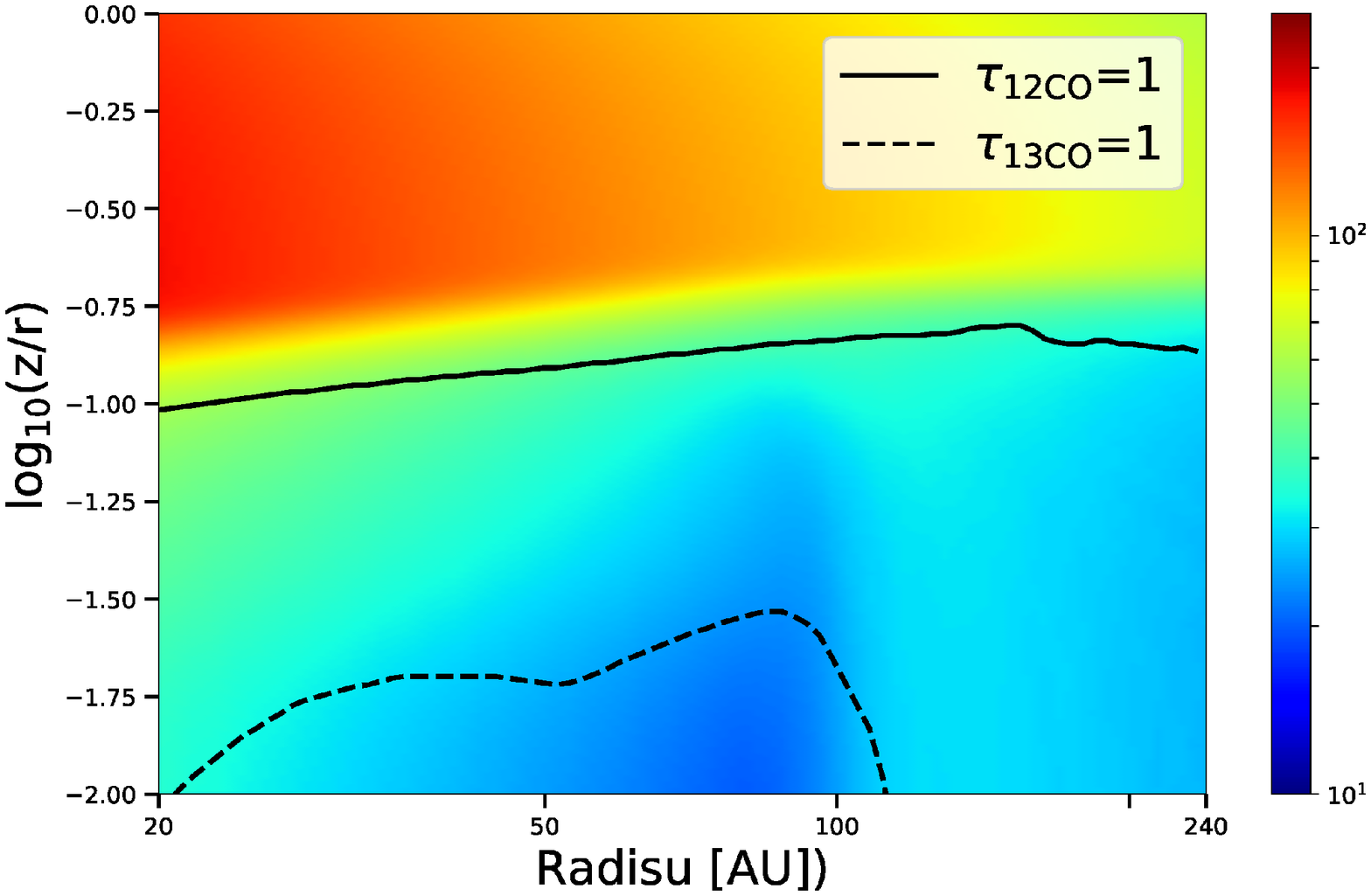}}
\subfigure[The T$_{\rm dust}$ distribution at the disk midplane]{\label{fig:temp}\includegraphics[width=0.45\textwidth]{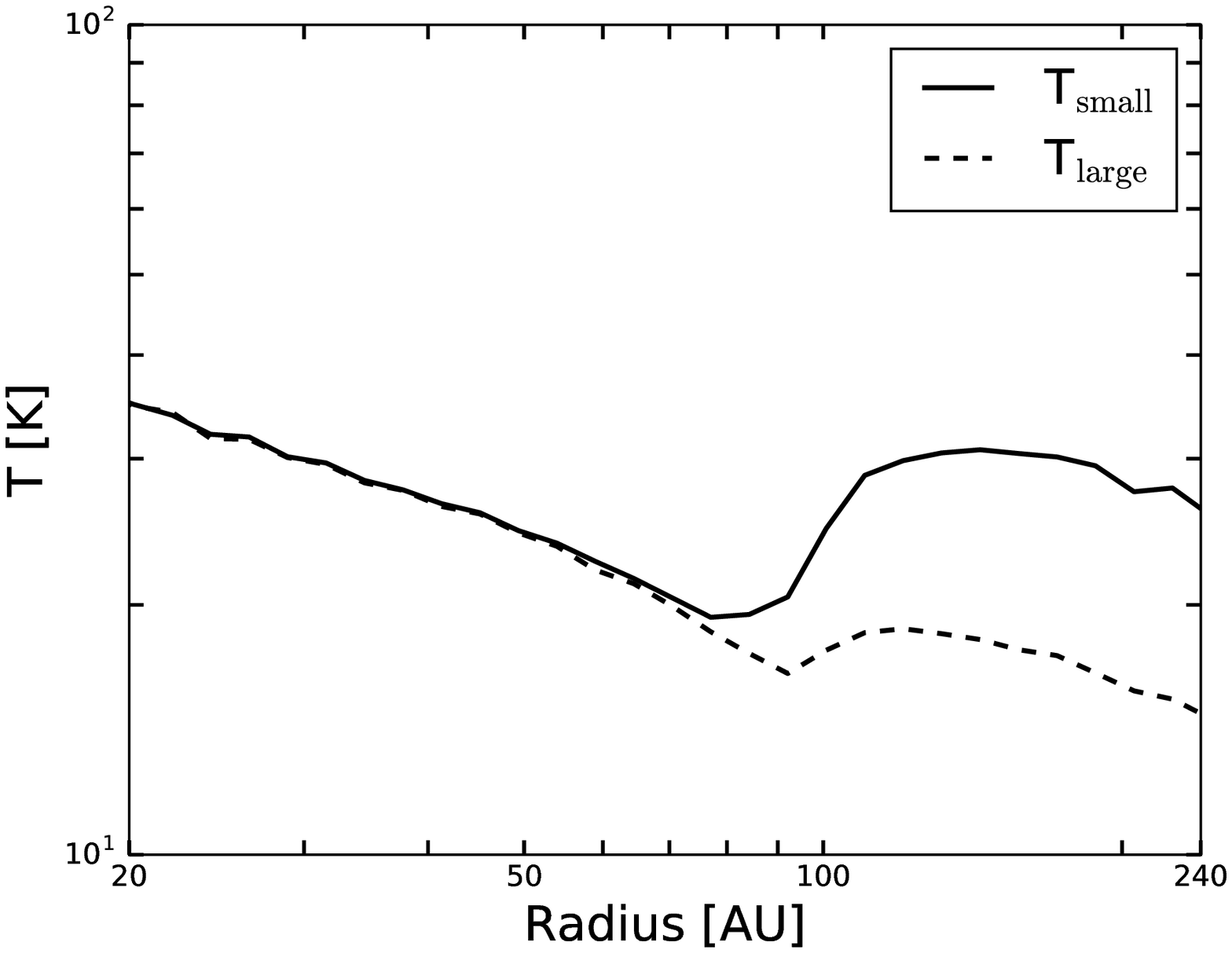}}
\subfigure[The T$_{\rm gas}$ distribution at the photosphere of the $^{12}$CO and $^{13}$CO lines]{\label{fig:t_tau1}\includegraphics[width=0.45\textwidth]{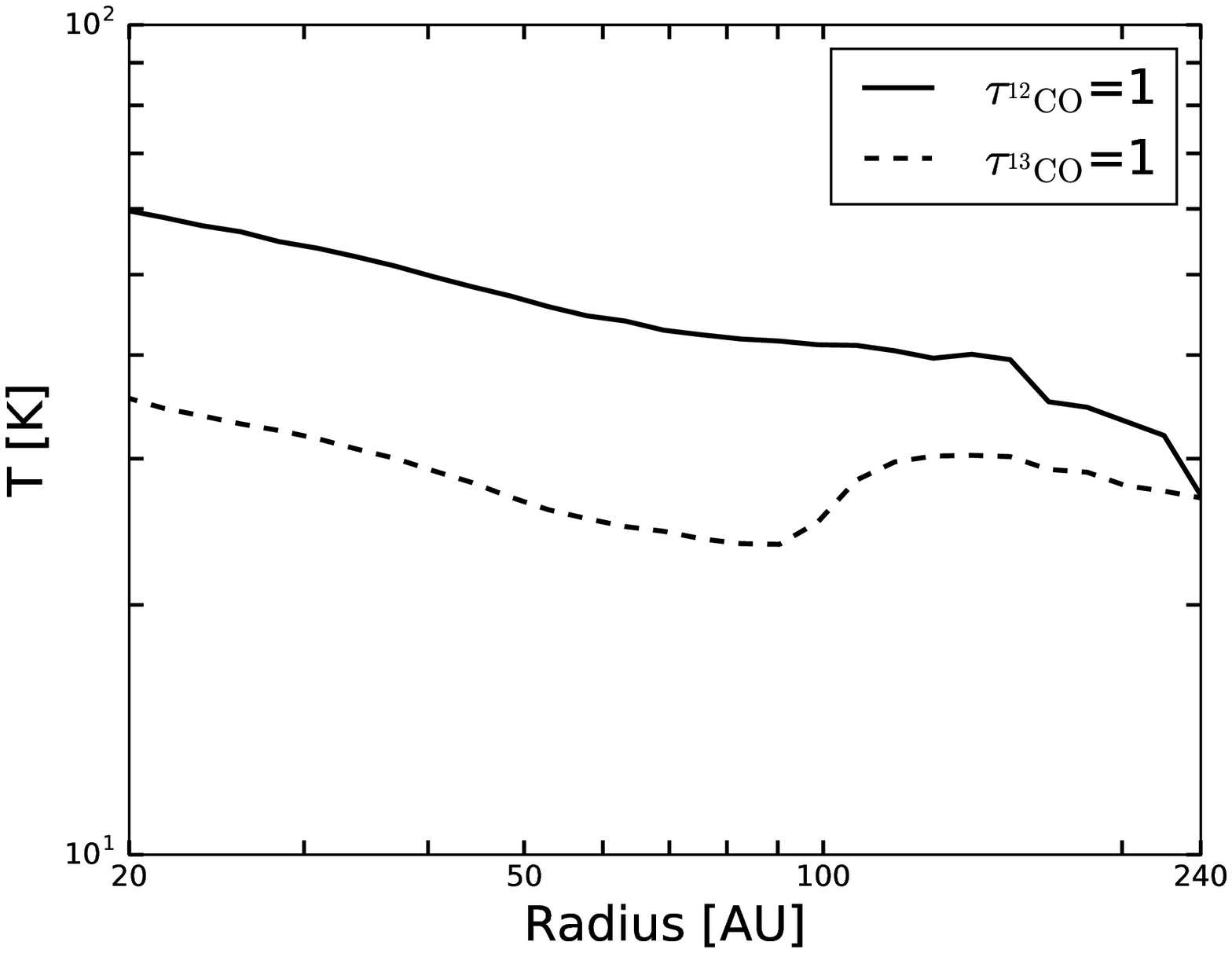}}
\subfiguretopcapfalse
\caption{ (a) The dust surface density profile (black) and C$^{18}$O column density (red) profile derived from our observations (solid) and the fitting results (dashed) of Equation (\ref{eq:eq3}) and (\ref{eq:NC18O}). 
(b) The gas temperature distribution in the disk $r-z$ plane derived by the radiative transfer calculation. 
The solid and dashed lines indicate the photospheres ($\tau_{\nu}=1$) of $^{12}$CO and $^{13}$CO emission lines, respectively.
(c) The temperature distribution of the small (solid) and large (dashed) dust grains at the disk midplane. The T$_{\rm small}$ distribution shows a bump beyond the dust gap at 90 au.
(d) The gas temperature distribution at the photosphere of $^{12}$CO (solid) and $^{13}$CO (dashed) emission lines shown in Figure \ref{fig:figure7}(b). The gas temperature bump around 130 au is seen at the photosphere of $^{13}$CO line, but no bump at the photosphere of $^{12}$CO line.  
}
\label{fig:figure7}
\end{figure*}

\subsubsection{Dust concentration on the gas pressure bump}
\label{subsubsec:dust trap model}

Here, we discuss another possible formation mechanism of the dust gap-ring structure in the outer region of the dust disk. 
We assume the situation that the dust disk becomes more compact than the gas disk through the radial drift of the dust grains in the disk. At the outer edge of the dust disk, the transition of optical depth occurs due to the steep decrease of dust surface density. Then, the temperature at the midplane increases by the direct irradiation heating from the hot surface layer of the disk. This temperature variation leads to the pressure bump formation just beyond the edge of the dust disk. If there is a small amount of dust grains in the outside of the dust disk, they drift inward and concentrate on the pressure bump so that the ring structure is formed. This mechanism is consistent with the observed features: a dust ring beyond the dust gap, more compact dust disk than the gas disk, and the temperature bump in $^{13}$CO line emission beyond the dust gap.

We perform the radiative transfer calculation to derive the temperature distributions in the disk $r-z$ plane when the dust disk is more compact than the gas disk by using a published code HO-CHUNK \cite[]{Whitney2003a,Whitney2003b,Whitney2004,Whitney2013}. We adopt the stellar mass of $\rm 1.657~M_{\odot}$, the effective temperature of 4800 K, and the stellar radius of $\rm 3.113~R_{\rm \odot}$ \cite[]{Villebrun2019}. We also use the dust surface density derived from the observed I$_{\rm cont}$ at $\rm T_{dust}=T_{ B,peak,^{13}CO}$. For the radiative transfer calculations, we use the fitted dust surface density profile within the radius of $\rm 20~au< r < 100~au$ to consider the compact dust disk. 
Although the dust temperature will change step by step during the radial drift of the grains in reality, here we assume a fixed dust surface density profile for the simplicity.
We obtain the fitted dust surface density profile of
\begin{equation}
\Sigma_{\rm d,RT} = A\left(\frac{r}{10\ {\rm au}}\right)^{B}\exp\left[-\left(\frac{r}{C\ {\rm au}}\right)^{10}\right]
\label{eq:eq3}
\end{equation}
where $A=0.27\pm0.02~\rm g~cm^{-2}$, $B=-0.45\pm0.06$, and $C=86.0\pm1.66$ au. It is presented as the black dashed line in Figure \ref{fig:sigma_d}. The dust distribution is effectively cut off at 100 au.

In our calculation, we introduce two dust components: small dust grains and large dust grains. The minimum and maximum sizes are 0.0025 ${\rm \mu}$m and 0.2 ${\rm \mu}$m for small dust grains and 0.01 ${\rm \mu}$m and 1 mm for large dust grains, respectively. The size distribution of both dust components is $n(a) \propto a^{-3.5}$, where $n(a)$ is the number density per unit radius and $a$ is the radius of dust grains. The opacity models for both small and large dust grains are described in \cite{Hashimoto2015} \cite[see also ][]{KimSH1994,Wood2002,Dong2012ApJ_750_161D}. We adopt Equation (\ref{eq:eq3}) as the surface density distribution of large dust grains and assume that the surface density of the small dust grains is 10 times smaller than that of the large dust \citep[$\Sigma_{\rm small}=0.1\Sigma_{\rm large}$; e.g.,][]{DAlessio2006}. We note that the small-to-large dust mass ratio for the size distribution of $n(a) \propto a^{-3.5}$ is 1\%. However, many practical calculations \citep[e.g.,][]{Andrews2011ApJ_732_42A, Dong2015gap} use the size distribution and change the small-to-large dust mass ratio independently because the results are not very different even if the mass ratio is assumed as 10\%. 
To obtain the density distributions of small and large dust grains, we set the scale height of them, 
\begin{equation}
 H_{\rm l} = \rm 2~au \left(\frac{r}{100~au}\right)^{1.25},
\end{equation}
\begin{equation}
 H_{\rm s} = \rm 6.5~au \left(\frac{r}{100~au}\right)^{1.25},
\end{equation}
where $H_{\rm l}$ and $H_{\rm s}$ are the scale height of large and small dust grains, respectively. We assume that the scale height of small dust grains is similar to the gas scale height. The scale height of the large dust grains is assumed to be about three times smaller than that of the small dust grains \footnote{If we assume that the gas surface density is $\rm 10~g~cm^{-2}$ and turbulence strength $\alpha = 10^{-2}$ at 70 au, we obtain that the scale height of 1 mm dust, which is the maximum size of the large dust, is about three times smaller than the gas scale height. In reality, this ratio will depend on the radius, but here we assume the constant ratio for simplicity. }. Using these scale heights, we evaluate the density of small and large dust grains, whose density distributions in $z$ direction are proportional to $\exp(-z^2/(2H_{\rm l,s}^2))$. We note that the dependence of model parameters on the resulting disk structures is out of the scope of this work and remains to be explored.

Figure \ref{fig:temp_rz} shows the temperature distribution in the disk $r-z$ plane generated by the radiative transfer calculation. We overlay the $\tau_{\nu} =1$ layer (i.e., photosphere) of $^{12}$CO and $^{13}$CO line emissions as the solid and dashed lines. To obtain the optical depth of the CO isotopologue lines, we evaluate the surface density of $^{12}$CO and $^{13}$CO from the C$^{18}$O column density obtained from the C$^{18}$O line observations, shown as the red solid line in Figure \ref{fig:sigma_d}. We fit the column density as follows,
\begin{equation}
N_{\rm C^{18}O} = A \left[1+\left( \frac{r}{B\ {\rm au}} \right)^{C}\right]^{-1},
\label{eq:NC18O}
\end{equation}
where $A=(4.46\pm0.04)\times10^{16}~{\rm cm^{-2}}$, $B=44.40\pm0.5$ au, and $C=2.12\pm0.03$. It is presented as the red dashed line in Figure \ref{fig:sigma_d}. We also assume the abundance ratio of $\rm^{12}CO/C^{18}O=444$ and $\rm ^{12}CO/^{13}CO=67$ \citep{Qi2011}. Since $^{12}$CO line emission is optically thick, the photosphere of $^{12}$CO line is seen at the hot upper surface of the disk. On the other hand, since $^{13}$CO line is marginally optically thick, the photosphere of $^{13}$CO line is located around or below the disk midplane. 

Figure \ref{fig:temp} shows the temperature distribution of dust grains at the disk midplane obtained from the radiative transfer calculations. The solid and dashed lines indicate the temperatures of small and large dust grains, respectively. The figure shows that T$_{\rm small}$ starts to increases at the outer edge of the dust gap at $\rm r\sim90~au$. In contrast, T$_{\rm large}$ does not increase much in the same region of the disk. 

The temperature distribution can be explained as follows. The temperature at the disk midplane is determined by the balance between the heating from the hot surface layer irradiated by the stellar radiation and cooling by the dust thermal emission. In the region $\rm r\lesssim 90$ au, the disk is optically thick against the irradiation from the hot surface layer so that the temperature of the large and small dust grains have the same temperature. On the other hand, since the disk is optically thin in the region $\rm r \gtrsim 90$ au, the small and large dust grains are directly irradiated by the hot disk surface. 
Since the wavelength dependence on the opacity of small dust grains is steeper than that of the large dust grains, the opacity of the small dust grains is higher for the irradiation from the hot disk surface ($\sim30~\mu$m) but lower for their thermal emission ($\sim100~\mu$m) than those of the large dust grains. Thus, the temperature of the small dust grains is higher than that of the large dust grains, resulting in the occurrence of the temperature bump in the region $\rm r\gtrsim90$ au. Since the gas temperature will be determined by the collisional heating with small dust grains, this temperature distribution causes the gas pressure bump around $\rm r\sim130$ au. If there is some amount of the dust in the region $\rm r \gtrsim 90$ au, they can be concentrated at this gas pressure bump and form the observed ring structure. Here, we assume that there is enough dust in the region $\rm r > 90$ au in addition to Equation (\ref{eq:eq3}) to make the ring structure as a result of the radial drift of the dust and to determine the gas temperature with the small dust temperature. We need further investigation of this scenario to justify this assumption.

Figure \ref{fig:t_tau1} shows the temperature distributions at the photosphere for $^{12}$CO line emission as solid line. There is no temperature bump beyond the dust gap at 90 au since it is at the hot upper disk surface. We also show the temperature distribution at the photosphere of the $^{13}$CO line emission as a dashed line, but we plot the midplane temperature if the photosphere of $^{13}$CO line is located below the midplane. Since the optical depth of $^{13}$CO line emission is smaller than that of $^{12}$CO line, it traces the temperature at the layer closer to the midplane so that there is a bump in temperature distribution in the region $\rm r\gtrsim 90$ au. This is consistent with the observed intensity bump of $^{13}$CO line around 130 au but nothing for the $^{12}$CO line. On the other hand, the C$^{18}$O line is optically thin and intensities in 20 K and 30 K are similar in LTE condition. Thus, we cannot see the bump in the C$^{18}$O line, too. We note that the optical depth of $^{13}$CO seems lower than the unity in the region of $\rm r>120$ au because Figure \ref{fig:temp} shows only the upper half-disk in the z-direction. When integrating the full vertical disk, the optical depth of $^{13}$CO becomes larger than unity up to $\rm r\sim160$ au but $\sim0.7$ in the region $\rm r>160$ au. Considering the large uncertainty of the line ratio in Figure \ref{fig:figure2}(b), it does not significantly conflict with our observations.


Meanwhile, the dust gap-ring structure in the CR Cha disk may be formed through different mechanisms. As shown in Figure \ref{fig:temp}, the dust temperature at the disk midplane inside the dust gap is $\sim20$ K. Since this is comparable to the CO sublimation temperature ($\sim15-25$ K), the dust gap and ring could be created by dust sintering effect \citep{Okuzumi2016}. 
Secular gravitational instability \citep[secular GI][]{Ward2000,Youdin2011,Michikoshi2012} is another possible mechanism of dust ring structure formation \citep{Takahashi2014}. As mentioned in Section \ref{subsec:dustgas}, the disk has small Q value and/or large dust-to-gas mass ratio. These features are suitable for the growth of the secular gravitational instability in the disk \citep[][]{Takahashi2014,Takahashi2016,Latter2017,Tominaga2019}.
Also, the existence of the dead zone is another possible cause to produce the dust gap-ring structure in the disk. If the gas gap is generated at the outer edge of the dead zone by viscous instability \citep[e.g.,][]{Flock2015,Hasegawa2015}, dust grains can be piled up there. The temperature bump at the outer edge of the gap is also generated by the heating of dust grains \citep[e.g.,][]{Hasegawa2010}.
To distinguish the dust gap-ring formation mechanisms, further investigations are required, for example, the size distribution of the dust particles, since the dust sintering mechanism indicates that the dust radius is small in a dust ring structure.

\section{Summary}
\label{sec:summary}

In this paper, we present the images of the dust continuum at 225 GHz and the CO isotopologue $J=2-1$ emission lines of the CR Cha disk observed by the ALMA. The observed dust continuum image shows a dust gap at $\rm r\sim90$ au and a faint dust ring at $\rm r\sim120$ au. We derive the radial profile of dust surface density to investigate the gap-ring structure by assuming the peak brightness temperature of $^{13}$CO emission line as dust temperature T$_{\rm dust}$. Using the Gaussian fitting, we find that the dust gap is located at the radius of 90 au and has the gap depth of $\sim20\%$ at the gap center with $\sim8$ au width on average.

We analyze $^{12}$CO, $^{13}$CO and C$^{18}$O $J=2-1$ emission lines to investigate the gas disk of CR Cha. The moment 8 maps, the channel maps, and the averaged spectra of the CO isotopologue lines inside the radius of 240 au show the absorption feature which might be caused by the Cha I foreground cloud at the velocity of $\sim4.5$ km s$^{-1}$. The azimuthally averaged radial profiles of the integrated CO isotopologue line intensities show that the gas disk is extended up to 240 au, which is much larger than the dust disk. The ratios between the CO isotopologue line emissions are larger than 0.2 at all radii, larger than the typical abundance ratios of CO isotopologue in the protoplanetary disks. It indicates that $^{12}$CO and $^{13}$CO lines are optically thick. To derive the gas temperature, we make the moment 8 maps of $^{12}$CO and $^{13}$CO lines without dust continuum subtraction because they are optically thick lines. The azimuthally averaged radial profile of the $^{13}$CO peak intensity shows a small bump around $\sim$130 au, which may indicate the temperature bump in the $^{13}$CO line emitting region. 

We investigate two possible mechanisms to reproduce the observed dust gap at 90 au, the dust ring structure at 120 au, and a bump of $^{13}$CO emission line at 130 au: gap opening by a planet and dust concentration on the gas pressure bump at the outer edge of the dust gap. 

For investigating the scenario of gap opening by a planet, we compare the observed surface density distribution around the gap to the model surface density obtained by numerical simulations done by \cite{Kanagawa2017}. We obtain the planet mass of $M_{\rm p}\approx0.99M_{\rm J}\left( \alpha/0.01\right)^{1/2}$, where $\alpha$ is the turbulent viscous parameter, can reproduce the observed dust surface density around the gap. However, this value has large uncertainties due to some reasons: the synthesized beam size comparable to the gap width only constrains the upper limit of the gap depth and the dust gap structure may not trace the gas gap structure. Thus, this value should be considered as a very rough estimation. 

For investigating the scenario of dust concentration on the gas pressure bump at the outer edge of the dust gap, we set the radiative transfer calculation with two dust components: small and large dust grains. When the dust disk becomes more compact than the gas disk by the radial drift of the dust grains in the disk, the steep gradient of dust surface density around the outer edge of the dust disk leads to the efficient heating by the irradiation from the hot upper disk surface to the small dust grains than the large dust grains. As a result, the gas temperature bump beyond the dust disk is produced because the small dust grains are coupled with the gas.
It can make the pressure bump and the dust grains may be concentrated at the pressure peak formed beyond the dust disk. Also, since the optical depth of $^{13}$CO line emission is smaller than that of $^{12}$CO, it traces the temperature around the midplane so that a bump of $^{13}$CO line is seen at 130 au, but not for $^{12}$CO line.
We note that the dust sintering, secular gravitational instability, and the existence of the dead zone at the disk midplane also can explain the observed dust gap-ring structure and gas temperature bump beyond the dust gap qualitatively. We need further investigation to develop quantitative interpretation in the future.

\acknowledgments
We would like to thank the referees for comments that improved this paper. 
This paper makes use of the following ALMA data: ADS/JAO.ALMA\#2017.1.00286.S. ALMA is a partnership of ESO (representing its member states), NSF (U.S.) and NINS (Japan), together with NRC (Canada) and NSC and ASIAA (Taiwan) and KASI (Republic of Korea), in cooperation with the Republic of Chile. The joint ALMA observatory is operated by ESO, AUI/NRAO, and NAOJ. 
Data analysis was carried out on the Multi-wavelength Data Analysis System operated by the Astronomy Data Center (ADC), National Astronomical Observatory of Japan. 
This work is supported by MEXT Grants-in-Aid for Scientific Research JP17H01103, JP18H05441, JP19K03910, and NAOJ ALMA Scientific Research Grant Numbers 2018-10B.
Y.H. is supported by the Jet Propulsion Laboratory, California Institute of Technology, under a contract with the National Aeronautics and Space Administration.

\software{HO-CHUNK \cite[]{Whitney2003a,Whitney2003b,Whitney2004,Whitney2013}, makemask (https://github.com/kevin-flaherty/ALMA-Disk-Code/blob/master/makemask.py), CASA \cite[v5.3;][]{McMullin2007ASPC_376_127M} }

\bibliographystyle{apj}
\bibliography{All_papers}

\clearpage

\appendix
\renewcommand\thefigure{\thesection.\arabic{figure}} 

\section{The channel maps of CO isotopologue lines}
\label{apsec:channel}

We present the channel maps of $^{12}$CO, $^{13}$CO and C$^{18}$O isotopologue lines to clearly show where the absorption feature is. The black contours overlaid in each panel indicate the Keplerian rotation model with the stellar mass 2 M$_{\odot}$, the disk inclination $31^{\circ}$, and disk position angle $36.2^{\circ}$. We used the python code\footnote{https://github.com/kevin-flaherty/ALMA-Disk-Code/blob/master/makemask.py} written by Kevin Flaherty for generating the Keplerian rotation model. At the velocity of $\rm V\sim4.5~km~s^{-1}$, absorption features appear. The Keplerian mask shows what regions are affected by this absorption in the moment 0 and 8 maps of the CO isotopologue lines (see Figure \ref{fig:figure1} and \ref{fig:figure4}).

\setcounter{figure}{0}    
\begin{figure*}[h]
\centering  
\includegraphics[width=0.83\textwidth]{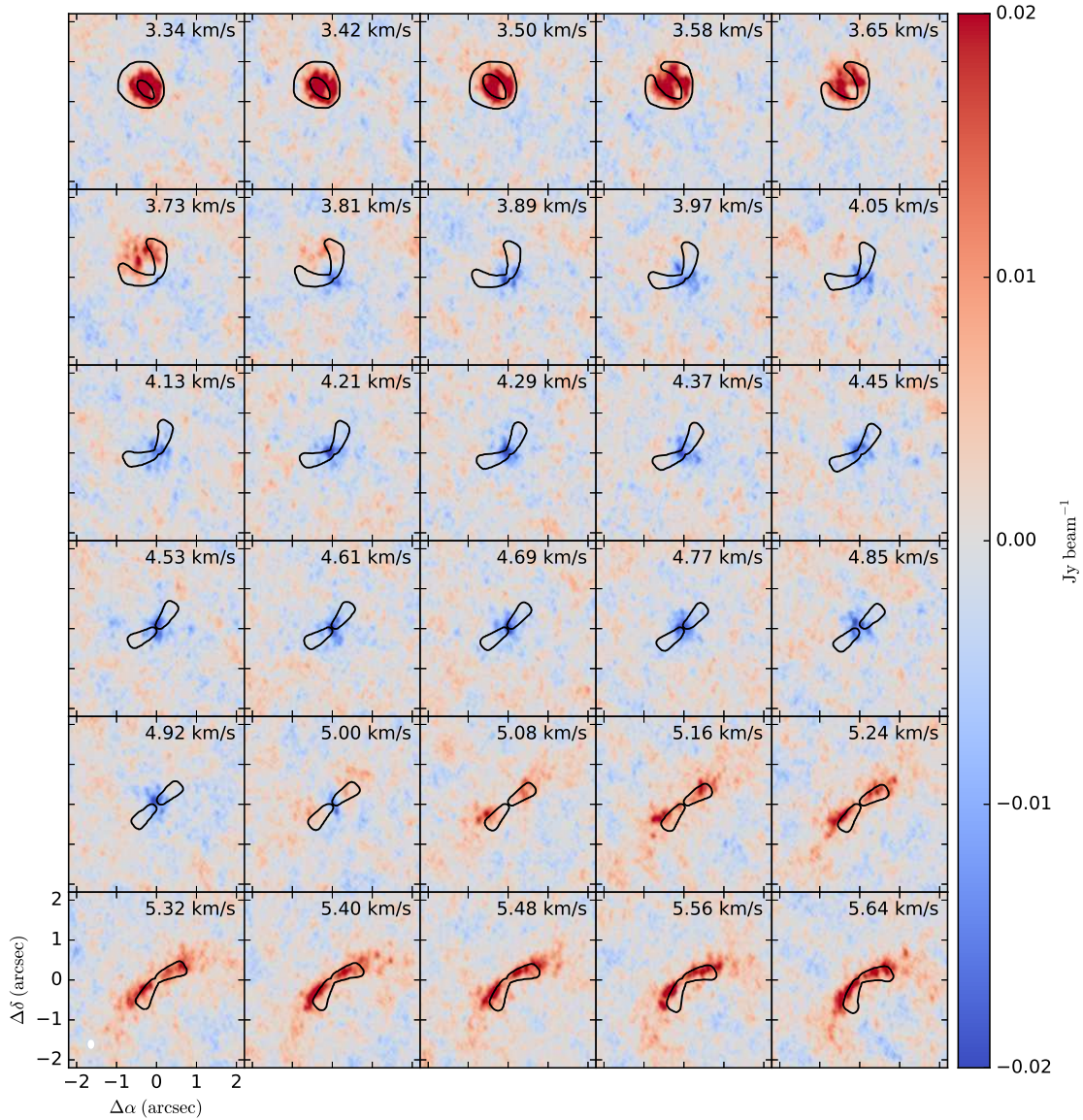}
\caption{ The $^{12}$CO $J=2-1$ line channel maps from 3.3 km s$^{-1}$ to 5.6 km s$^{-1}$. The channels are showing the absorption feature in the velocity of $\rm 3.8<v<5.1~km~s^{-1}$. The black contours indicate the Keplerian rotation model derived by the 31$^{\circ}$ disk inclination and 36.2$^{\circ}$ disk position angle. }
\label{apfig:figurea1}
\end{figure*}

\begin{figure*}[p]
\centering  
\includegraphics[width=0.83\textwidth]{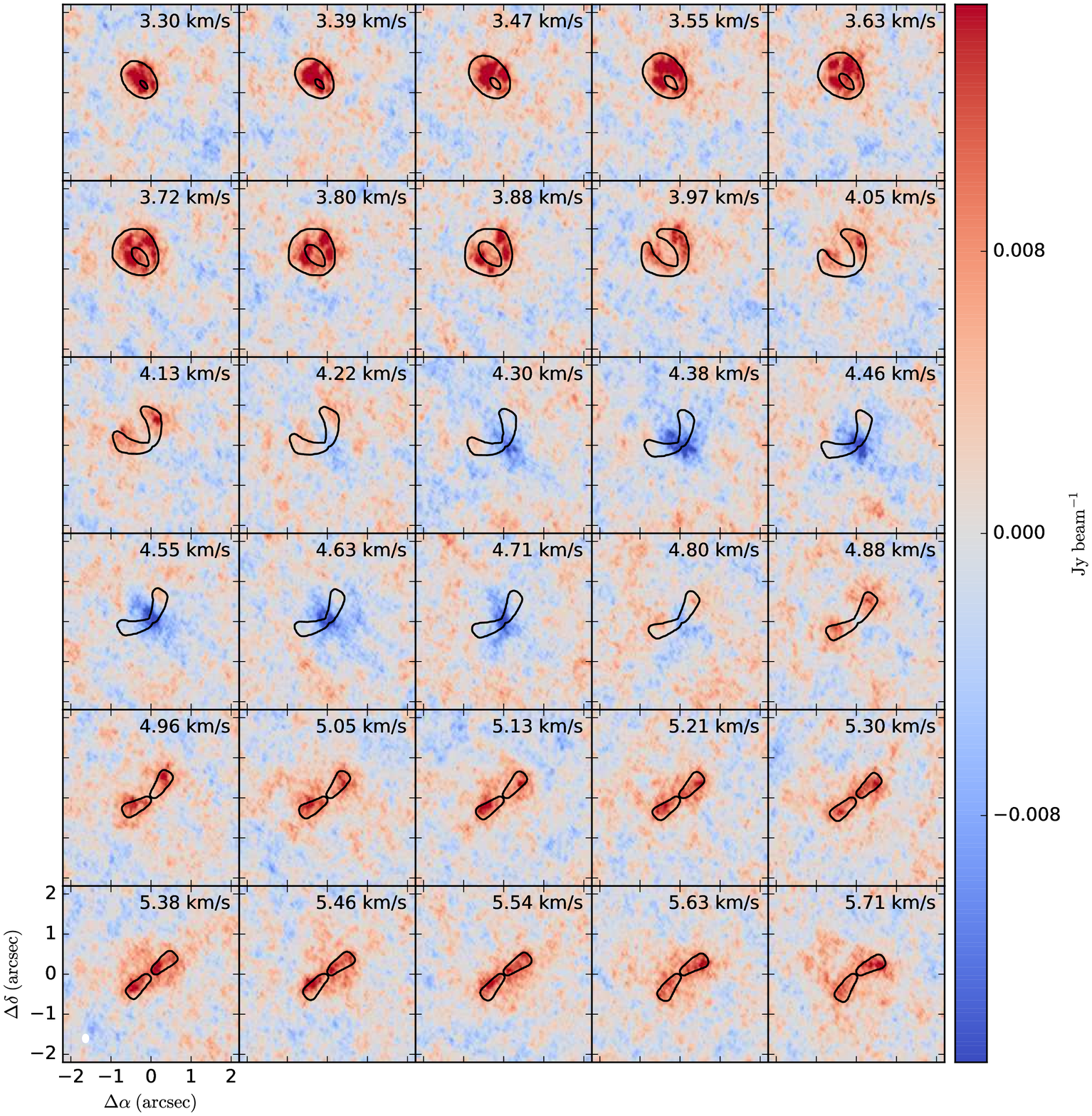}
\caption{ The $^{13}$CO $J=2-1$ channel maps from 3.3 km s$^{-1}$ to 5.7 km s$^{-1}$. The channels are showing the absorption feature in the velocity of $\rm 4.3<v<4.8~km~s^{-1}$. The black contours indicate the Keplerian rotation model derived by the 31$^{\circ}$ disk inclination and 36.2$^{\circ}$ disk position angle.}
\label{apfig:figurea2}
\end{figure*}

\begin{figure*}[p]
\centering  
\includegraphics[width=0.83\textwidth]{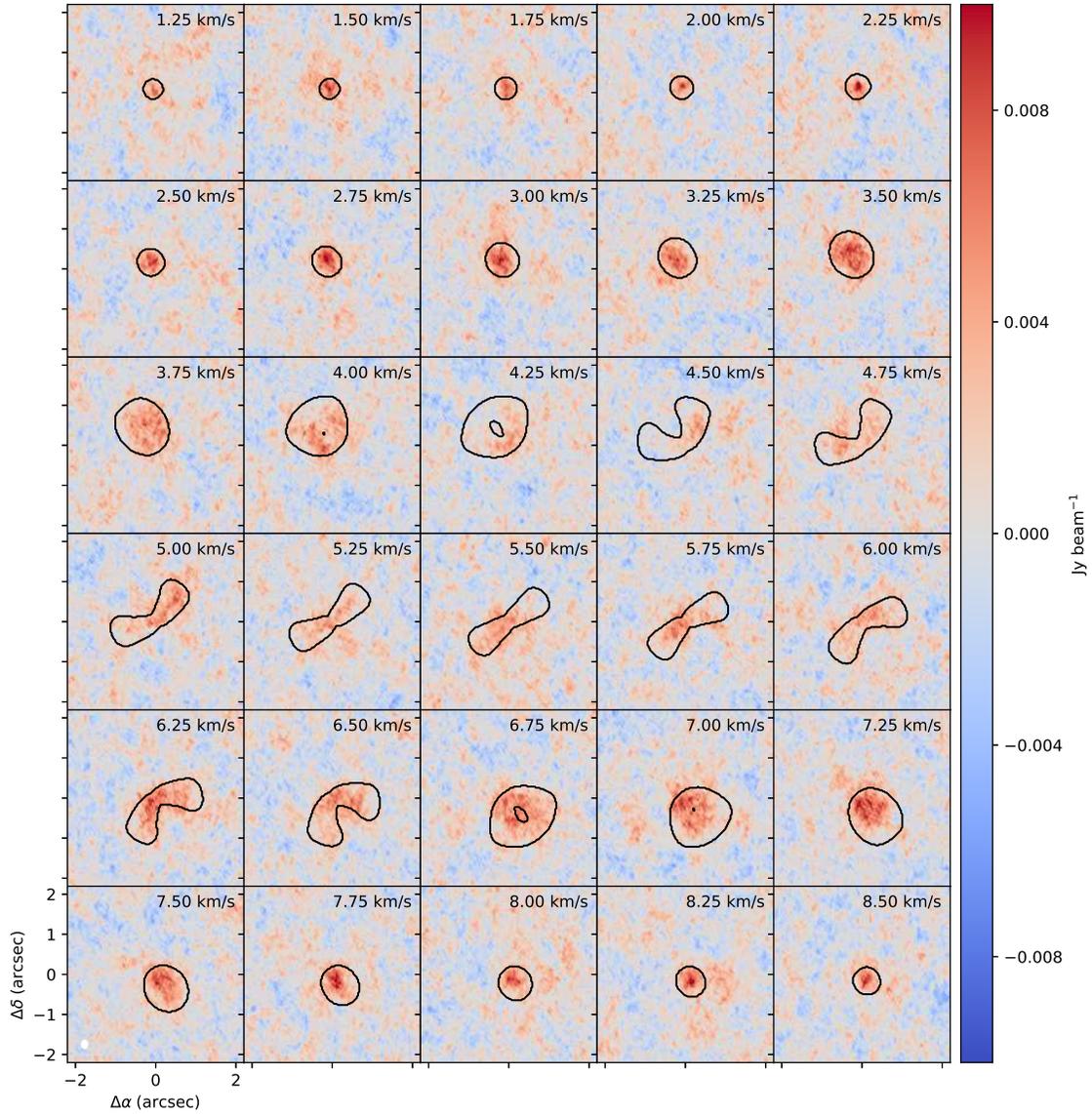}
\caption{ The C$^{18}$O $J=2-1$ channel maps from 1.25 km s$^{-1}$ to 8.5 km s$^{-1}$ by smoothing the velocity resolution from 0.08 km s$^{-1}$ to 0.25 km s$^{-1}$ for increasing the signals. There is no clear absorption feature in the velocity of $\rm 4<v<5~km~s^{-1}$. The black contours indicate the Keplerian rotation model derived by the 31$^{\circ}$ disk inclination and 36.2$^{\circ}$ disk position angle.}
\label{apfig:figurea2}
\end{figure*}
\end{document}